\documentclass[pra,letterpaper,10pt,twocolumn]{revtex4}
\usepackage{graphicx}
\usepackage[final,dvips]{epsfig}
\usepackage{color}
\usepackage{dcolumn}  % Align table columns on decimal point
\usepackage{bm}       % bold math
\usepackage{array}    % table making
\usepackage{amssymb,amsmath}
\usepackage[colorlinks,citecolor=red]{hyperref}
\emergencystretch=\maxdimen
\include{MyCommand}

\begin{document}

\title{Discovering Phases, Phase Transitions and Crossovers
through Unsupervised Machine Learning: A critical
examination}

\author{Wenjian Hu$^{1,2}$}
\author{Rajiv R.P. Singh$^{1}$}
\author{Richard T. Scalettar$^{1}$}
\affiliation{
$^1$Department of Physics, University of California Davis, 95616 CA  USA\\
$^2$Department of Computer Science, University of California Davis,
95616 CA  USA }

\begin{abstract} 
We apply unsupervised machine learning techniques, mainly principal component analysis (PCA), to compare and
contrast the phase behavior and phase transitions in several classical
spin models - the square and triangular-lattice Ising models, the
Blume-Capel model, a highly degenerate biquadratic-exchange spin-one
Ising (BSI) model, and the 2D XY model, and examine critically what
machine learning is teaching us.  We find that quantified principal
components from PCA not only allow
exploration of different phases and symmetry-breaking, but can
distinguish phase transition types and locate critical points. We show
that the corresponding weight vectors have a clear physical
interpretation, which is particularly interesting in the frustrated
models such as the triangular antiferromagnet, where they can point to
incipient orders.  Unlike the other well-studied models, the properties
of the BSI model are less well known.  Using both PCA and conventional
Monte Carlo analysis, we demonstrate that the BSI model shows an absence
of phase transition and macroscopic ground-state degeneracy.  The
failure to capture the `charge' correlations (vorticity) in the BSI
model (XY model) from raw spin configurations points to some of the
limitations of PCA.  Finally, we employ a nonlinear unsupervised machine
learning procedure, the `antoencoder method', and demonstrate that it
too can be trained to capture phase transitions and critical points.
\end{abstract}

\maketitle

%%%%%%%%%%%%%%%%%%%%%%%%%%%%%%%%%%%%%%%%%%%%%%%%%%%%%%%%%%%%%%%%%%%%%%%%%%
\section{Introduction}
%%%%%%%%%%%%%%%%%%%%%%%%%%%%%%%%%%%%%%%%%%%%%%%%%%%%%%%%%%%%%%%%%%%%%%%%%%

In the age of big data, machine learning has become an indispensable
tool whose utility transcends scientific and academic
boundaries\cite{Witten}. From social networking\cite{Nowell, Kyumin} to
object and image recognition\cite{Bishop, Rosten}, from advertising to
finance\cite{Pang}, from engineering to medicine\cite{King}, from
biological physics\cite{McKinney, Schafer} to
astrophysics\cite{VanderPlas}, wherever there is preponderance of
information and real data, machine learning is helping to find and
quantify patterns and even discover basic laws\cite{Crutchfield}.

Condensed Matter Physics is a relatively late arrival in this field.
Although isolated applications of machine-learning have been made over
many years, it is only recently that a concerted effort towards the use
of these methods for addressing problems in many-body physics has
started to emerge\cite{Carleo, Broecker, Torlai, Carrasquilla, Wang,
Schoenholz, Wei, Torlai2, Nieuwenburg, Khatami}. On the one hand, machine learning algorithms,
such as deep learning\cite{Hinton2, Krizhevsky}, have profound
connections to the foundations of statistical physics\cite{Engel,
Mehta}.  Scaling and renormalization\cite{Cardy} are core principles
that underlie our ability to make sense of macroscopic phenomena, in
particular, their simplicity and universality despite incredible
microscopic complexity. It is perhaps not surprising that the way
forward for machines to learn from large data sets would incorporate
similar principles.

On the other hand, machine learning is also being used, mostly in
conjunction with Monte Carlo\cite{Wolff, Robert, Katzgraber} data, to recognize
phases and phase-transitions\cite{Carrasquilla, Broecker, Wang, Nieuwenburg, Khatami, Wetzel}, to
maximize accuracy of variational Monte Carlo methods in quantum
many-body systems\cite{Carleo}, to guide choices of Monte Carlo 
moves \cite{Walter, Graham}, to
explore overcoming the famous sign-problem bottleneck\cite{Broecker}, and
to infer spectral functions by performing analytic continuation from
imaginary-time to real-time data \cite{Arsenault}. Both supervised
learning\cite{Carrasquilla, Liu_Junwei}, which depends on first training
the system with trial data-sets and unsupervised learning\cite{Wang},
which works with no prior training, have been found useful.

In this paper, we focus further on Monte Carlo simulations and examine
the extent to which unsupervised machine learning can succeed in
deciphering and distinguishing different physics. We introduce a simple
biquadratic-exchange spin-one Ising model that has a macroscopic
ground-state degeneracy which can be lifted by a small bilinear-exchange
term. We contrast its behavior with more well understood and studied
cases-- the Ising ferromagnet on a square lattice\cite{Binder_Ising} and Ising
antiferromagnet on a triangular lattice\cite{Wannier, Stephenson1,
Stephenson2}, the Blume-Capel
model\cite{Blume, Capel, BEG, Jain, Liu} (which has both first and
second order transitions to phases with true long range order), and the
two-dimensional XY model\cite{Barouch, Kosterlitz}, which has a low
temperature power-law correlated phase.

Our primary results are: (a) Simple machine learning methods are good at
recognizing order and symmetry breaking and the temperature region of
sharpest change. Thus, it is possible to locate the transition
temperature with reasonable accuracy.  (b) Strong first-order
transitions can be easily distinguished from second-order transitions by
the evolution of the principal component distribution.  (c) A crossover
can be distinguished from a phase-transition. (d) Principal components
are related to Fourier modes and order parameters and their size
dependence can be used to extract critical exponents.  (e) The principal
components are particularly interesting in the fully frustrated
triangular antiferromagnet, where they point to incipient orders upon
perturbation which are not evident at first glance.  
%(f) Sometimes
%machine learning can promote variation in local quantities, such as the
%density, as a greater discriminant than emergent correlations.  
(f) Finally, the analysis is sensitive to the nature of the data provided to
the machine-learning algorithms. The use of bare spin-configurations
versus bond, triangle or square-plaquette spin, or local vorticity
configurations provide complementary inferences and are not readily
discovered from the other.

In the conclusion, we will discuss why automated machine learning can be
an important part of our toolbox for studying phase transitions using
large scale numerical simulations.

%%%%%%%%%%%%%%%%%%%%%%%%%%%%%%%%%%%%%%%%%%%%%%%%%%%%%%%%%%%%%%%%%%%%%%%%%%
\section{Models}
%%%%%%%%%%%%%%%%%%%%%%%%%%%%%%%%%%%%%%%%%%%%%%%%%%%%%%%%%%%%%%%%%%%%%%%%%%

%% \subsection{Classical Ising Model}

In this paper, we investigate several classical models of 
phase transitions.  The simplest is the two-dimensional
square-lattice Ising model:
\begin{align}
H &= -J\sum_{\left\langle i,j \right\rangle} S_iS_j
\label{eq:ham_Ising}
\end{align}
where the `spin' $S_i=\pm 1$.  Positive $J$ corresponds to
ferromagnetism, where the energy favors aligned spins, while for
negative $J$ antiferromagnetic configurations have lower energy.  On a
bipartite lattice with nearest neighbor interactions, 
%such as the case considered here, 
the thermodynamics of $J>0$ and $J<0$ are identical and, for the square
lattice, a phase transition to a magnetically ordered state occurs as
the temperature $T$ is reduced below the critical value $T_c/J = 2/\ln(1
+ \sqrt{2}) \approx 2.269$\cite{Onsager}.  We also study the
antiferromagnetic triangular lattice Ising model (TLIM) as an example of
a fully frustrated model with no finite temperature phase transition and
macroscopic ground state degeneracy.

%% \subsection{Blume Capel Model}

The Blume-Capel model (BCM) is a generalization of
Eq.~\ref{eq:ham_Ising} which allows three values $S_i=\pm 1,0$ with
associated energy, 
\begin{align} H &=
-J\sum_{\left\langle i,j \right\rangle} S_iS_j + 
%% \sum_i \Delta^{\phantom{2}}_i S^2_i
\Delta \sum_i S^2_i
\label{eq:ham_Blume_Capel} 
\end{align} 
Again we choose the coupling $J=+1$ between nearest neighbor spins on a
square lattice with periodic boundary conditions.  Qualitatively
speaking, the BCM allows `site vacancies' ($S_i=0$) in addition to
non-zero magnetic moments ($S_i=\pm 1$).  The parameter $\Delta$ of
Eq.~\ref{eq:ham_Blume_Capel} controls the vacancy density. For $\Delta
\rightarrow -\infty$, $S_i = 0$ is energetically unfavorable, and the
BCM reduces to the Ising model.  As $\Delta$ increases, the second order
Ising-like transition from ferromagnet to paramagnet becomes first order
above a tricritical point $(T/J,\Delta/J)=(0.609(4),1.965(5))$
\cite{Plascak98} and occurs with a discontinuous jump in magnetization
and vacancy density.  The BCM provides a description of systems ranging
from metamagnets and ternary alloys, to multicomponent spins  and the
dynamics of rough surfaces
\cite{BEG,MukamelBlume74,Lajzerowicz75,Newman83,Zahraouy04,Brito07}.  It
has also been studied for nonuniform $\Delta_i$ \cite{Pittman08} to
understand situations where the spin density varies across the lattice.

%% \subsection{Biquadratic-exchange Spin-one Ising Model}

The BCM was further generalized by Blume, Emery, and Griffiths to
\begin{align}
H &= -J \sum_{\left\langle i,j \right\rangle} S_i S_j + K \sum_{\left\langle i,j \right\rangle} S_i^2 S_j^2 + \sum_i\Delta_i S_i^2  
\label{eq:ham_BEG}
\end{align}
to incorporate an additional biquadratic interaction $K
\sum_{\left\langle i,j \right\rangle} S_i^2 S_j^2$,  and used to study
$^3$He-$^4$He mixtures \cite{BEG}.  A variant of Eq.~\ref{eq:ham_BEG},
which we call the biquadratic-exchange spin-one Ising (BSI) model, 
\begin{align}
H &= -J \sum_{\langle\langle i,k \rangle\rangle} S_i S_k 
+ K \sum_{\left\langle i,j \right\rangle} S_i^2 S_j^2,
\label{eq:ham_BSI}
\end{align}
will be studied here.  As with the BCM and BEG model, $S_i$ takes
values: $+1$, $0$ or $-1$.  Our primary interest is in the purely
biquadratic limit with $J=0$.  Because it has been less well studied, we
discuss in this section some of the basic physics of the BSI model
revealed by conventional Monte Carlo before turning to the results of
machine learning approaches.

\begin{figure}[!h]
\includegraphics[width=0.98\columnwidth]{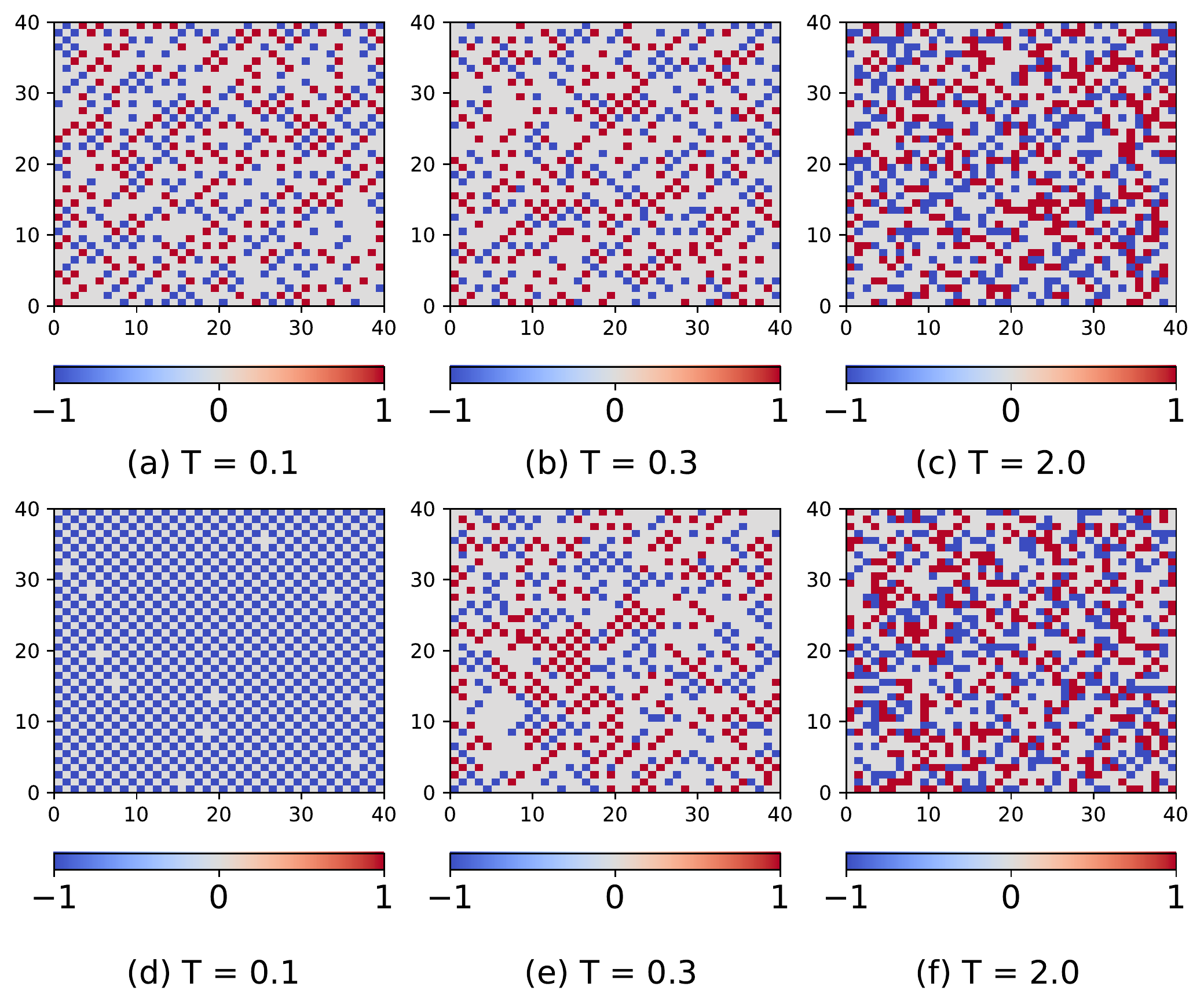}  
\caption{
\underline{Top row:} Spin configuration snapshots from conventional
Metropolis Monte Carlo simulations of the BSI model with fixed $K=1$,
$J=0$ and linear lattice size $L=40$.  At high temperatures there is a
random mix of the three spin possibilities $S_i = 0,\pm 1$.  As $T$
decreases below $K$, sites occupied by $S_i = \pm 1$ are gradually surrounded by
near neighbor vacancy sites $S_i=0$.  \underline{Bottom row:} Similar
spin configuration snapshots with $J=0.1$.  The evolution from $T=2.0$
(panel f) to $T=0.3$ (panel e) is similar to that at $J=0.0$ (panel c to
panel b).  However, in this case an ordered phase emerges (panel d) at
the lowest $T=0.1$.  In all BSI calculations reported in this paper $K$
is fixed to $1$ as the energy scale.  
\label{fig:Modified_Ising_Model_1}
}
\end{figure}

Since $K>0$, it is energetically unfavorable to have adjacent sites of
the BSI model both `occupied' with $S_i=\pm 1$.  Therefore, as $T$ is
lowered one expects all occupied sites to be surrounded by vacancies.
As long as $J=0$, there is no preference for one $S_i=\pm1$ orientation
over the other. 
If one sublattice is occupied with $S_i=0$, the other
sublattice is free to take any spin value. 
Thus the $J=0$ BSI model has
macroscopic ground-state entropy. A question arises: Is there a phase
transition to a checkerboard `charge-ordered' phase where the $S_i=\pm
1$ variables preferentially occupy one sublattice?  Thus, along with the
triangular-antiferromagnet, this model provides machine learning with
the challenge of dealing with high degeneracy and deciding between the
presence and absence of a phase transition.

Simulations of this $J=0$ case Fig.~\ref{fig:Modified_Ising_Model_1}
panels (a,b,c) can be contrasted with a second-neighbor bilinear
exchange interaction $J=0.1$, where conventional Monte Carlo analysis
reveals a sharp second order phase transition at temperature
$T/K=0.163$.  Fig.~\ref{fig:Modified_Ising_Model_1} panels (d,e,f) show
spin configuration snapshots that go from the ordered ferromagnetic
phase at $T=0.1$ to an intermediate degenerate regime at $T=0.3$, to a
high temperature phase at $T=2.0$. 

Finally, the XY model in two dimensions,
\begin{align}
H = -J \sum_{\langle ij \rangle} 
{\rm cos}\,\big(\,\theta_i - \theta_j \,\big)
\label{eq:ham_XY}
\end{align}
in which sites are occupied by a continuous spin $\theta_i \in (-\pi,\pi]$, offers a situation in which the performance of machine learning
for a Kosterlitz-Thouless (KT) \cite{Barouch, Kosterlitz} phase
transition between exponentially decaying spin correlations and a line
of critical points with power law decaying correlations can be evaluated
when provided with various sorts of real space snapshots - the spin
directions themselves or measures of local vorticity.

%%%%%%%%%%%%%%%%%%%%%%%%%%%%%%%%%%%%%%%%%%%%%%%%%%%%%%%%%%%%%%%%%%%%%%%%%%
\section{Methods of Machine Learning}
%%%%%%%%%%%%%%%%%%%%%%%%%%%%%%%%%%%%%%%%%%%%%%%%%%%%%%%%%%%%%%%%%%%%%%%%%%

Broadly speaking, machine learning can be divided into `unsupervised'
and `supervised' approaches.  In the latter case, a neural network, for
example, can be `trained' by providing a set of patterns along with a
desired `answer' (eg in the case of statistical mechanics whether the
pattern is ordered or not).  Thus the trained (supervised) network is
able to `generalize' and distinguish order from disorder on patterns to
which it had not been previously exposed.  In the former case, patterns
are given to the analysis procedure, but no `answers' are provided.  In
this paper, we utilize two methods for unsupervised machine learning,
`Principal Component Analysis' and the `Autoencoder' approach. These two
approaches have been described in \cite{Pearson, Wikipedia, Jolliffe,
Bourlard, Hinton, Ruslan}.  We provide a brief review here, with further
detail in the Appendix.

%%%%%%%%%%%%%%%%%%%%%%%%%%%%%%%%%%%%%%%%%%%%%%%%%%%%%%%%%%%%%%%%%%%%
\subsection{Principal Component Analysis} 
%%%%%%%%%%%%%%%%%%%%%%%%%%%%%%%%%%%%%%%%%%%%%%%%%%%%%%%%%%%%%%%%%%%%

Principal Component Analysis (PCA) \cite{Pearson, Wikipedia, Jolliffe} is a
strikingly simple method. In its implementation for our statistical
models, one constructs a matrix $\mathbf{S}$,
%% =[{\mathbf{S}}^T_{1}, \dots, {\mathbf  {S}}^T_{M}]^T$, 
each row of which is 
%% $\mathbf{S}_i=[s_1, \dots, s_N]$, 
a snapshot of the instantaneous values of the degrees of freedom from
a Monte Carlo simulation, {\it e.g.}~a listing of spins on each lattice site.
We denote by $M$ the total number of such configurations, and
therefore the number of rows of $\mathbf{S}$. It is convenient to compute the mean value 
$m_j=(1/M)\sum_{i}S_{ij}$, 
of each column, and
subtract that value from the entries in the column to
obtain the `centered data matrix' 
$\mathbf{X}$ with `column-wise zero empirical mean'. 
%% $\mathbf{X}=[{\mathbf{x}}^T_{1}, \dots,
%% {\mathbf  {x}}^T_{M}]^T=[{\mathbf{S}}^T_{1}-\mathbf{m}^T, \dots,
%% {\mathbf  {S}}^T_{M}-\mathbf{m}^T]^T$. 
To study phase transitions using
PCA, we choose $t$ evenly separated temperatures (or some other parameter) 
%in a range chosen to cross a phase transition 
and for each, we generate $n$ uncorrelated spin configurations
using a Monte Carlo simulation. 
The dimension of $\mathbf{X}$ is $M=nt$.

PCA then extracts features from the data in $\mathbf{X}$ by performing
an orthogonal transformation.  The result is the conversion of a set of
configurations of possibly correlated variables into a set of values of
linearly uncorrelated variables, named principal components, such that
the greatest variance of the data comes to lie on the first principal
component, the second greatest variance on the second principal
component, and so on\cite{Wikipedia}. Each succeeding component in turn
has the largest variance possible under the constraint that it is
orthogonal to the preceding components and the resulting vectors are an
uncorrelated orthogonal basis set. 

To put this more formally,
given the centered data matrix $\mathbf{X}$, the
transformation is defined by iteratively constructing
a set of $N$-dimensional `weight vectors'.
%% of weights $\mathbf{w}_{k}=[w_{1k},\dots ,w_{Nk}]^T$. 
The first is found by
%% weight vector $\mathbf{w}_{1}$ can be found by:
\begin{align}
{\mathbf  {w}}_{{1}}={\underset  {\Vert {\mathbf  {w}}\Vert
=1}{\operatorname {\arg \,max}}}\,\left\{\sum
_{i}\left(\mathbf{x}_i\cdot {\mathbf  {w}}\right)^{2}\right\}
\label{eq:pca_math}
\end{align}
where the arguments of the maxima (abbreviated to arg max) are the points of the domain of some function at which the function values are maximized. 
Succeeding weight vectors are obtained by subtracting the already
calculated principal components from $\mathbf{X}$ and repeating
Eq.~\ref{eq:pca_math}. 
The full principal components decomposition of
$\mathbf{S}$ can therefore be given as $\mathbf{P} = \mathbf{S}
\mathbf{W}$, where $\mathbf{W}=[{\mathbf{w}}_{1}, \dots, {\mathbf
{w}}_{N}]$. 

The inner products of the spin configuration $\mathbf{S}_i$, 
with the weight vectors $\mathbf{w}_j$ are termed its components $p_{ij}$,
\begin{align}
p_{ij}=\mathbf{S}_i \cdot \mathbf{w}_j
\label{eq:pca_math_3}
\end{align}
The `quantified principal components' are defined as the averages: 
\begin{align}
\left\langle \, |p_{j}|  \, \right\rangle=\frac{1}{n}\sum_i |p_{ij}|
\label{eq:pca_math_4}
\end{align}
where the average is taken over some appropriate subset of all
samples.  In our statistical mechanics application of PCA, it is natural
to do the quantification by averaging over all $n$ spin configurations with
the same temperature.  The central power of PCA is in cases where the
nature of the system is well described by a small number of `principal
components', e.g.~$p_{i1}$ and $p_{i2}$ only, and their associated
averages.

The principal components transformation can, equivalently, be thought
of in terms of
another matrix factorization method, the singular value decomposition
(SVD). It can be shown that the PCA weight vectors are eigenvectors of the
$N\times N$ real symmetric matrix $\mathbf{X}^T \mathbf{X}$, determined
by\cite{Jolliffe, White}: 
\begin{align} \mathbf{X}^T \mathbf{X}
\mathbf{w}_{n} = \lambda_n \mathbf{w}_{n} 
\label{eq:pca_math_2}
\end{align} 
Eigenvalues returned from Eq.~\ref{eq:pca_math_2}, sorted in
a descending order $\lambda_1\ge \lambda_2 \dots \ge \lambda_N \ge 0$,
represent variances of the input matrix $\mathbf{X}$ along corresponding
weight vectors.  
%It is conventional to denote the normalized eigenvalues
It is convenient to denote the normalized eigenvalues
$\tilde{\lambda}_n = \lambda_n/\sum_{i=1}^N \lambda_i$ as relative
variances.

%%%%%%%%%%%%%%%%%%%%%%%%%%%%%%%%%%%%%%%%%%%%%%%%%%%%%%%%%%%%%%%%
\subsection{The Autoencoder Approach}
%%%%%%%%%%%%%%%%%%%%%%%%%%%%%%%%%%%%%%%%%%%%%%%%%%%%%%%%%%%%%%%%

PCA has been shown to be effective in extracting aspects of
phase transitions\cite{Wang, Bradde}.
As discussed above, however, PCA is based on 
linear transformations of the input data.
Thus, it is natural to explore other, non-linear transformation,
methods.

In this spirit, we have also examined the use of an
`autoencoder'\cite{Bourlard, Hinton, Ruslan}, {\it i.e.}~an artificial
neural network used for unsupervised learning of efficient encodings by
exploiting the discovery of compressed representations.  The basic idea
is to pass information, in our case spin configurations, through an
intermediate layer whose number of `neurons' is considerably smaller
than the number of bits needed to encode the original, full spin
configuration.  The goal is to reproduce the original spin configuration
as best possible despite the constraint imposed by the reduced
information storage capacity of the hidden layer.  By attempting this
accurate reconstruction, the network may discover important
structure/patterns in the input data.

\begin{figure}[!h]
\includegraphics[width=0.6\columnwidth]{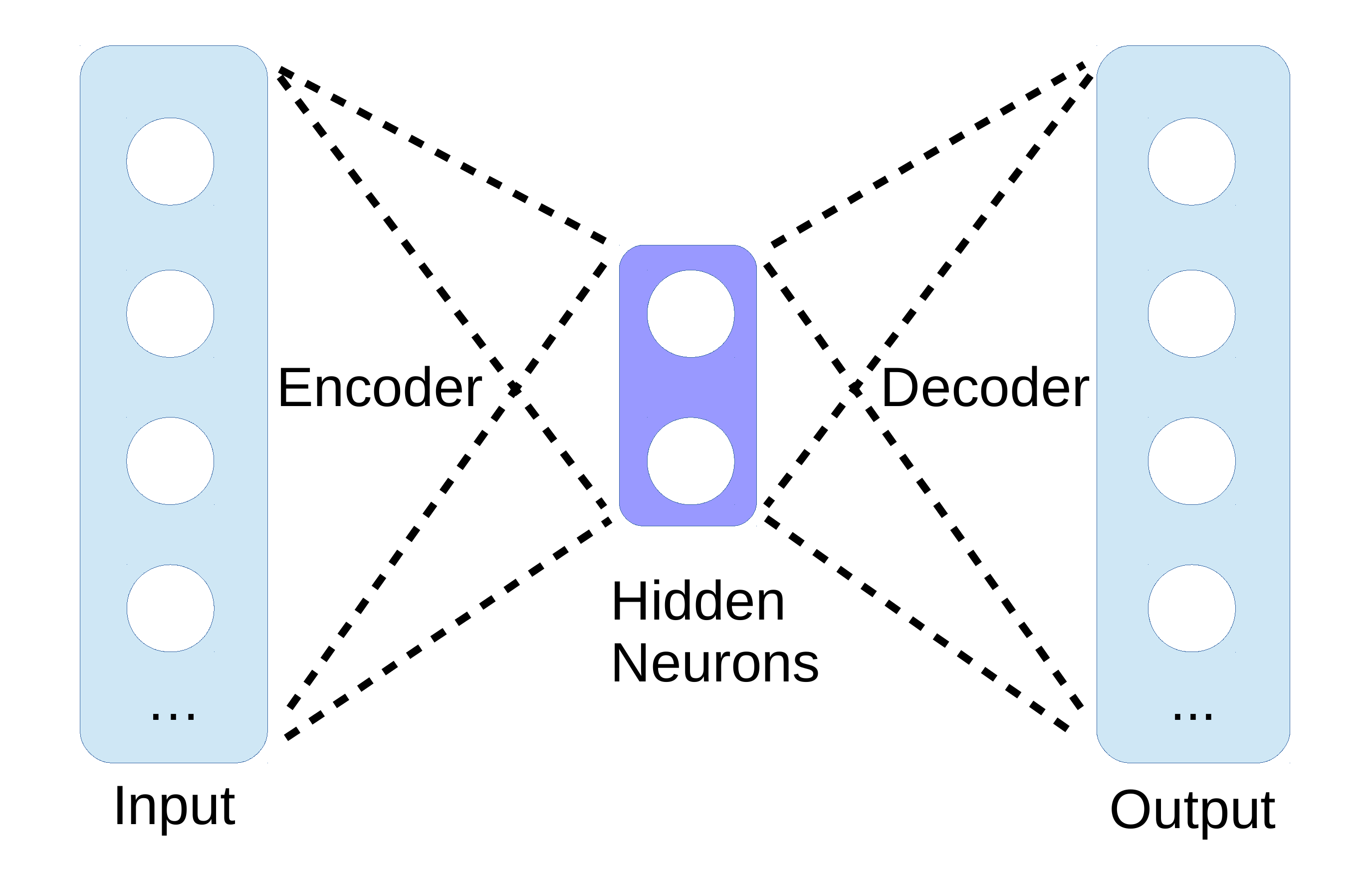}
\caption{ Schematic structure of an autoencoder with 2 hidden neurons.
CNNs are treated as the encoder and the decoder in this work.
\label{fig:autoencoding_1} } \end{figure}

A schematic autoencoder structure with two hidden neurons is shown in
Fig.~\ref{fig:autoencoding_1}. Although the autoencoder and PCA both use
the idea of dimensional reduction (in the case of the PCA by focussing
only on the one or two largest (`principal') components of the
decomposition in Eq.~\ref{eq:pca_math_2}) the autoencoder approach is
generally more powerful and more flexible than PCA.  Rather than being
restricted to linear transformations of the input data into principal
components, the autoencoder can incorporate nonlinear representations.
Given a set of spin configurations $\{\mathbf{S}_i\}$, in an autoencoder
the weights in the neural network can be trained through, for example,
backpropagation\cite{Witten, Rumelhart}, to return target values equal
to the inputs. In this work, we choose convolutional neural networks
(CNNs)\cite{Krizhevsky, Lawrence, Schmidhuber} as the encoder and the
decoder networks. As an alternative, artificial neural network based variational autoencoders can also be applied to tackle the problem, as reported in Ref.~\cite{Wetzel}.
 
%%%%%%%%%%%%%%%%%%%%%%%%%%%%%%%%%%%%%%%%%%%%%%%%%%%%%%%%%%%%%%%%%%%%%%%%%%
\section{Results and Analysis}
%%%%%%%%%%%%%%%%%%%%%%%%%%%%%%%%%%%%%%%%%%%%%%%%%%%%%%%%%%%%%%%%%%%%%%%%%%

%%%%%%%%%%%%%%%%%%%%%%%%%%%%%%%%%%%%%%%%%%%%%%%%%%%%%%%%%%%%%%%%%%%%%%%%%%
\subsection{PCA results of the Ising model}
%%%%%%%%%%%%%%%%%%%%%%%%%%%%%%%%%%%%%%%%%%%%%%%%%%%%%%%%%%%%%%%%%%%%%%%%%%

\begin{figure}[!h]
\includegraphics[width=0.98\columnwidth]{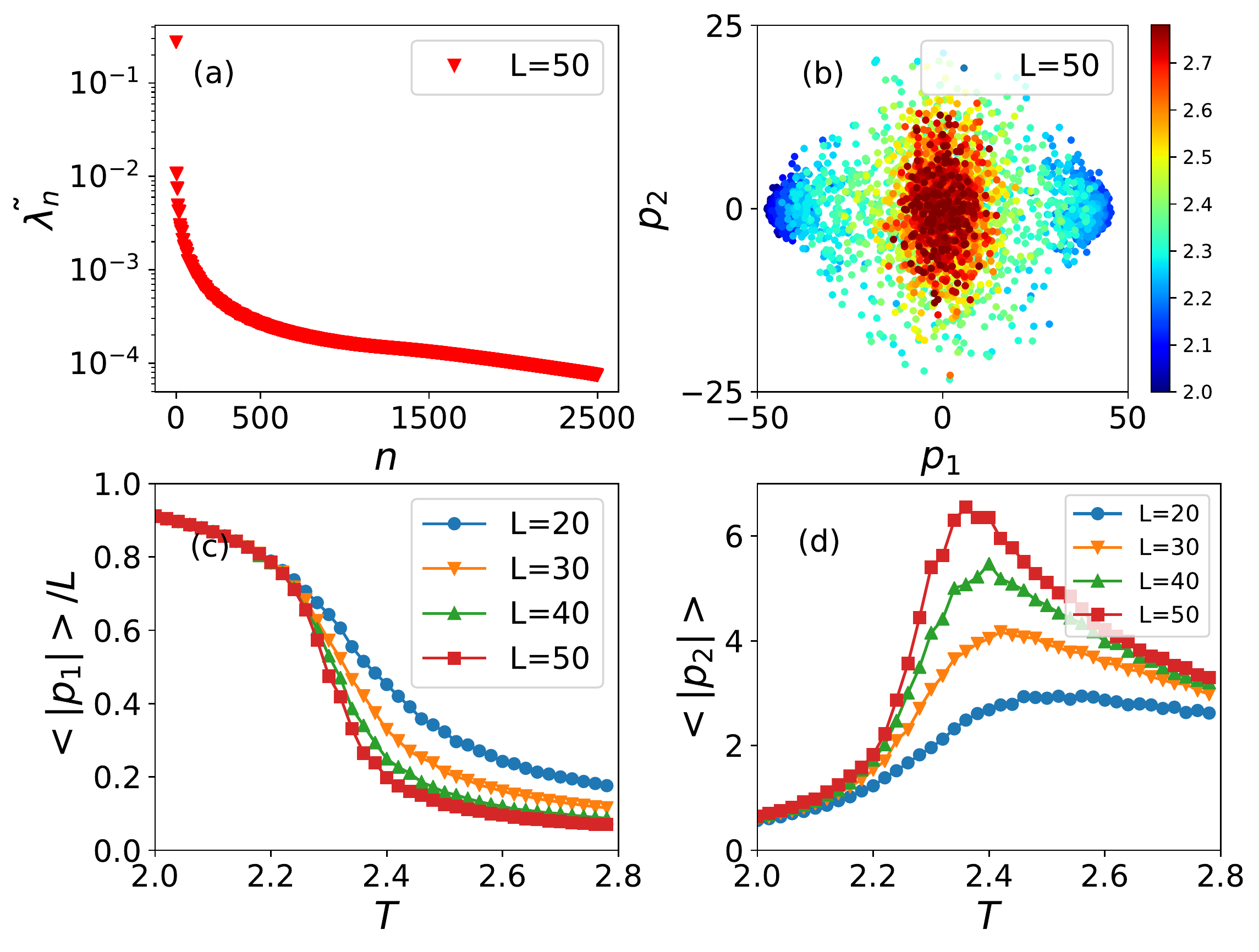}  
\caption{
PCA results for the 2D square lattice Ising model. (a) Relative variances
$\tilde{\lambda}_n$
obtained from the raw Ising configurations, with the horizontal axis indicating
corresponding component labels. (b) Projection of the raw Ising
configurations onto the plane of the two leading principal components.
The color bar on the right indicates the
temperature in units of $J$. (c) The normalized
quantified first leading component as a function of temperature. (d) The
quantified second leading component as a function of temperature. 
%In (c,d) the components are also normalized to lattice size $L$.
%% Peaks provide direct estimates of transition temperatures.
\label{fig:Regular_Ising_Model}
}
\end{figure}

By feeding a set of raw Ising spin configurations $\{\mathbf{S}_i\}$
into PCA, we generate results for the square-lattice Ising model, shown in
Fig.~\ref{fig:Regular_Ising_Model}.  
We have chosen $t=40$
temperatures in the range from $2.0$ to $2.8$ with $\Delta T = 0.02$.
These bracket the critical temperature $T_c\approx 2.269$.
For each temperature, we generate $n=10,000$ uncorrelated spin
configurations, so that $M=400,000$ configurations are used for
calculating weight vectors and relative variances. 
In Fig.~\ref{fig:Regular_Ising_Model} (a), a single dominant
principal component is observed;  
the largest $\lambda_k$, obtained through
Eq.~\ref{eq:pca_math_2},
is more than an order of magnitude greater than the next largest.
As we shall further argue below, the physical implication is that
there is a single
dominant spin pattern for the Ising model, corresponding to alignment
of all spins ({\it i.e.} ferromagnetism). 

Projecting raw spin configurations into the 2D plane spanned by the
first two principal components $p_{i1}$ and $p_{i2}$, we obtain
Fig.~\ref{fig:Regular_Ising_Model} (b).  For clarity of presentation,
only $100$ scatter points for each temperature are shown.
We see that above the critical temperature
the points lie in a single blob at the origin, but that
below $T_c$ a separation arises into two distinct regions.

In Fig.~\ref{fig:Regular_Ising_Model} (c), we plot the quantified first
leading component $\left\langle|p_{1}| \right\rangle/L$ versus
temperature, normalized to the system size $L$.  This is seen to mimic
the behavior associated with a measurement of the (absolute value of)
magnetization from a conventional Monte Carlo analysis.
Figure~\ref{fig:Regular_Ising_Model} (d) displays the quantified second
leading component $\left\langle|p_{2}| \right\rangle$ versus
temperature.  Its development is reminiscent of that of the
susceptibility $\chi$.

\begin{figure}[!h]
\includegraphics[width=0.98\columnwidth]{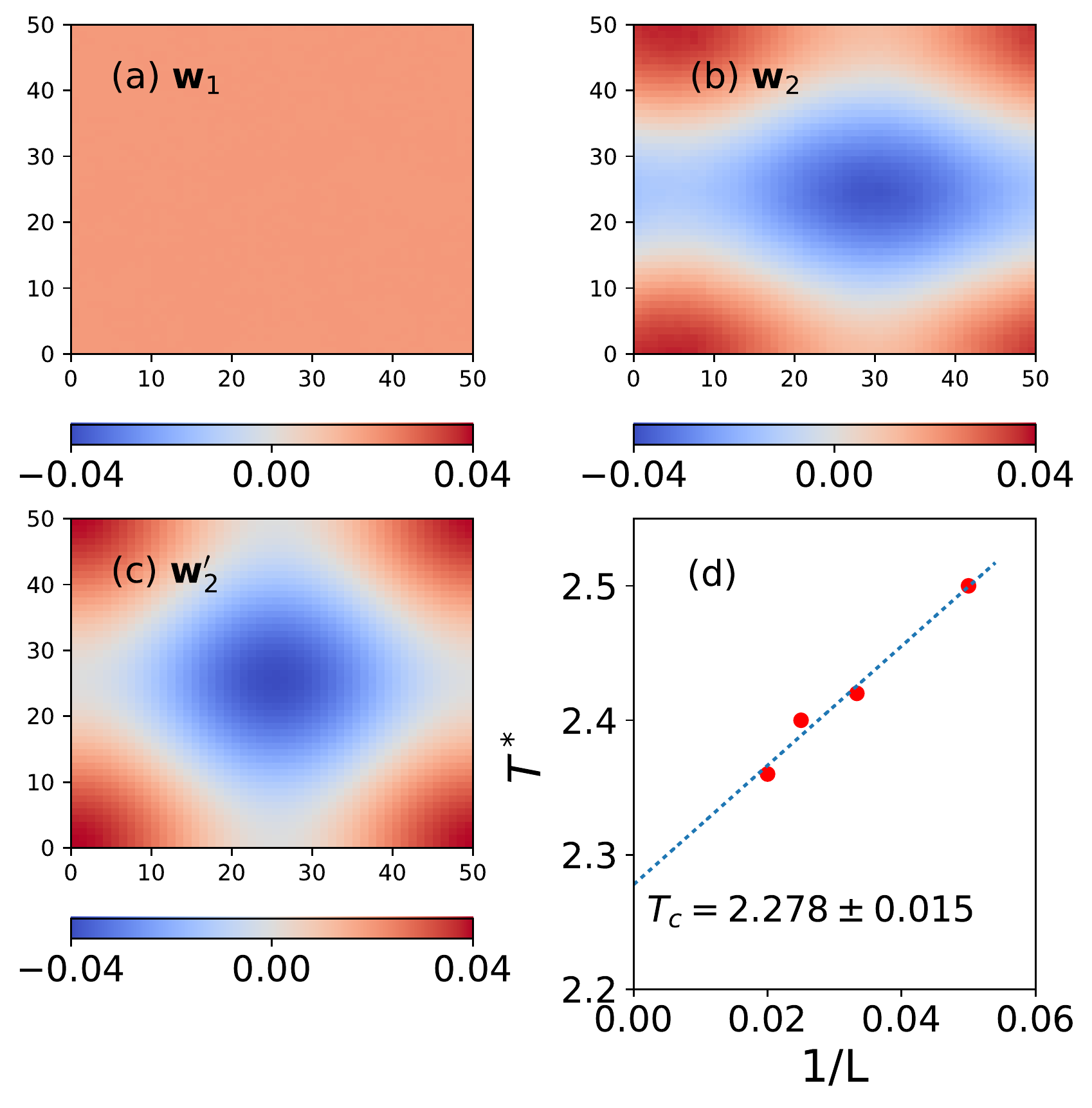} 
\caption{
(a) Visualization of the weight vector corresponding to the first leading
component.  (b) Visualization of the weight vector corresponding
to the second leading component.  
(c) The vector $\mathbf{w}^{\prime}_2$ defined by
Eq.~\ref{eq:pca_math_5}.
In (a-c) the lattice size $L=50$. 
(d) Peaks ($T^*$)
returned from the quantified second leading component as a function of
the inverse of the lattice dimension ($1/L$). The extrapolation was
performed using a linear least-squares fit. 
\label{fig:Regular_Ising_Model_weights}
}
\end{figure}

The weight vector corresponding to the first leading component is shown
in Fig.~\ref{fig:Regular_Ising_Model_weights} (a).  Ignoring
statistical/numerical variations, this weight vector is seen to be
roughly constant: $\mathbf{w}_1 = \frac{1}{L}[1,\dots,1]$.  Since the
first principal component $p_{i1}$ is the inner product of
$\mathbf{w}_1$ with $\mathbf{S}_i$ (Eq.~\ref{eq:pca_math}) the
connection to magnetization is clear.  The ability of an automatic
machine learning algorithm to recognize this order parameter is not
surprising as the order parameter of the model is so evident.  We have
checked that studying the variation of this with size at the critical
temperature $T_c$ can be related to known critical exponents for this 2D
Ising system. This is also not a surprise given that this principal
component is just the order parameter of the system.

Similarly, the weight vector corresponding to the second leading component is
shown in Fig.~\ref{fig:Regular_Ising_Model_weights} (b). To
understand the physical meaning of this weight vector, we compare it
in Fig.~\ref{fig:Regular_Ising_Model_weights} (c) with
\begin{align}
{\mathbf{w}}^{\prime}_{2} = \frac{1}{L}[\cos(\mathbf{r}_1\mathbf{k}_1), \dots, \cos(\mathbf{r}_N\mathbf{k}_1)] \nonumber \\
+\frac{1}{L}[\cos(\mathbf{r}_1\mathbf{k}_2), \dots, \cos(\mathbf{r}_N\mathbf{k}_2)]
\label{eq:pca_math_5}
\end{align}
where $\mathbf{r}_i$
%% =(r_{ix},r_{iy})$, 
are the lattice sites and
$\mathbf{k}_1=(0, \frac{2\pi}{L})$ and $\mathbf{k}_2=(\frac{2\pi}{L}, 0)$,
are the two Fourier wave vectors closest to the origin
$\mathbf{k}_0=(0,0)$.
There is clear similarity between panels (b,c).

We conclude that, for the ferromagnetic Ising model, PCA is building up
weight vectors corresponding to the Fourier modes of the spin
configuration.  In its ordered phase, the physics of the Ising model is
dominated by the single point $\mathbf{k}_0=(0,0)$, and hence PCA
reveals a single dominant eigenvalue.  Subleading eigenvalues are
associated with the next most ordered arrangements, {\it i.e.}~with a
single horizontal or vertical domain wall.  When supplied with
configurations of the ferromagnetic Ising model in the zero
magnetization ensemble, PCA generates four large eigenvalues
corresponding to horizontal or vertical domain walls\cite{Wang}.

Finally, Fig.~\ref{fig:Regular_Ising_Model_weights} (d), shows the peaks
($T^*$) returned from the quantified second leading component, versus
the inverse of the lattice dimension ($1/L$).  A linear finite size
scaling fit to these peaks yields $T_c \sim 2.278 \pm 0.015$, which
agrees reasonably well with the exact result $T_c/J \approx 2.269$.

%%%%%%%%%%%%%%%%%%%%%%%%%%%%%%%%%%%%%%%%%%%%%%%%%%%%%%%%%%%%%%%%%%%%
\subsection{PCA results of the Blume-Capel model}
%%%%%%%%%%%%%%%%%%%%%%%%%%%%%%%%%%%%%%%%%%%%%%%%%%%%%%%%%%%%%%%%%%%%

\begin{figure}[!h]
\includegraphics[width=0.98\columnwidth]{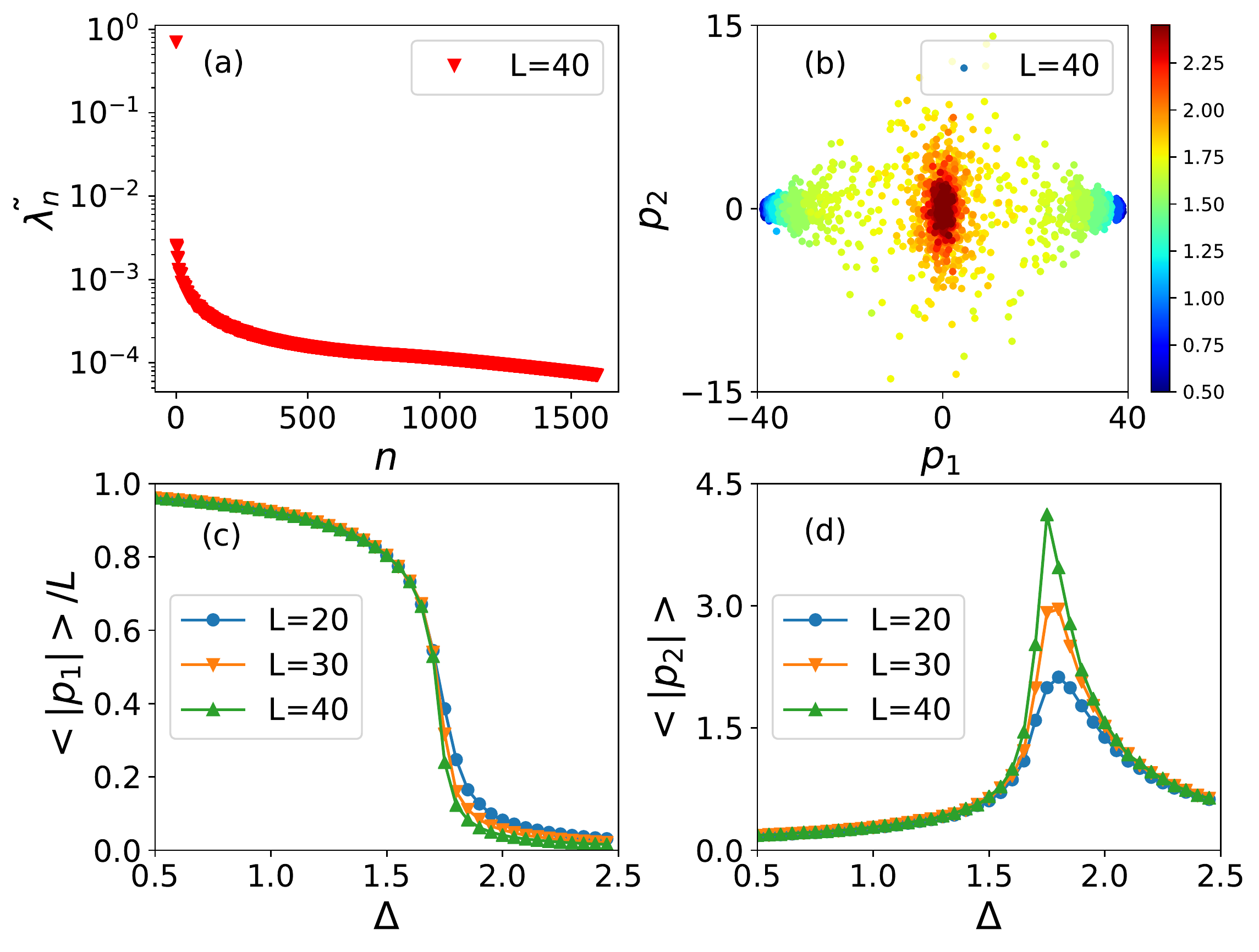}  
\caption{
PCA results for the BCM at fixed $J=1.0, T=1.0$ 
and sweeping $\Delta$. 
(a) Relative variances obtained from the raw Blume-Capel
configurations, with the horizontal axis 
indicating the corresponding component label. (b) Projection
of the raw Blume-Capel configurations onto the plane of leading two
principal components. The color bar on
the right indicates $\Delta$ in units of $J$. (c) The normalized
quantified first leading component as a function of
$\Delta$. (d) The quantified second leading component as a function
$\Delta$. Peaks provide estimates of the transition point $\Delta_c$. 
%In (c,d) the components are also normalized to lattice size $L$.
\label{fig:Blume_Capel_T10}
}
\end{figure}

\begin{figure}[!h]
\includegraphics[width=0.98\columnwidth]{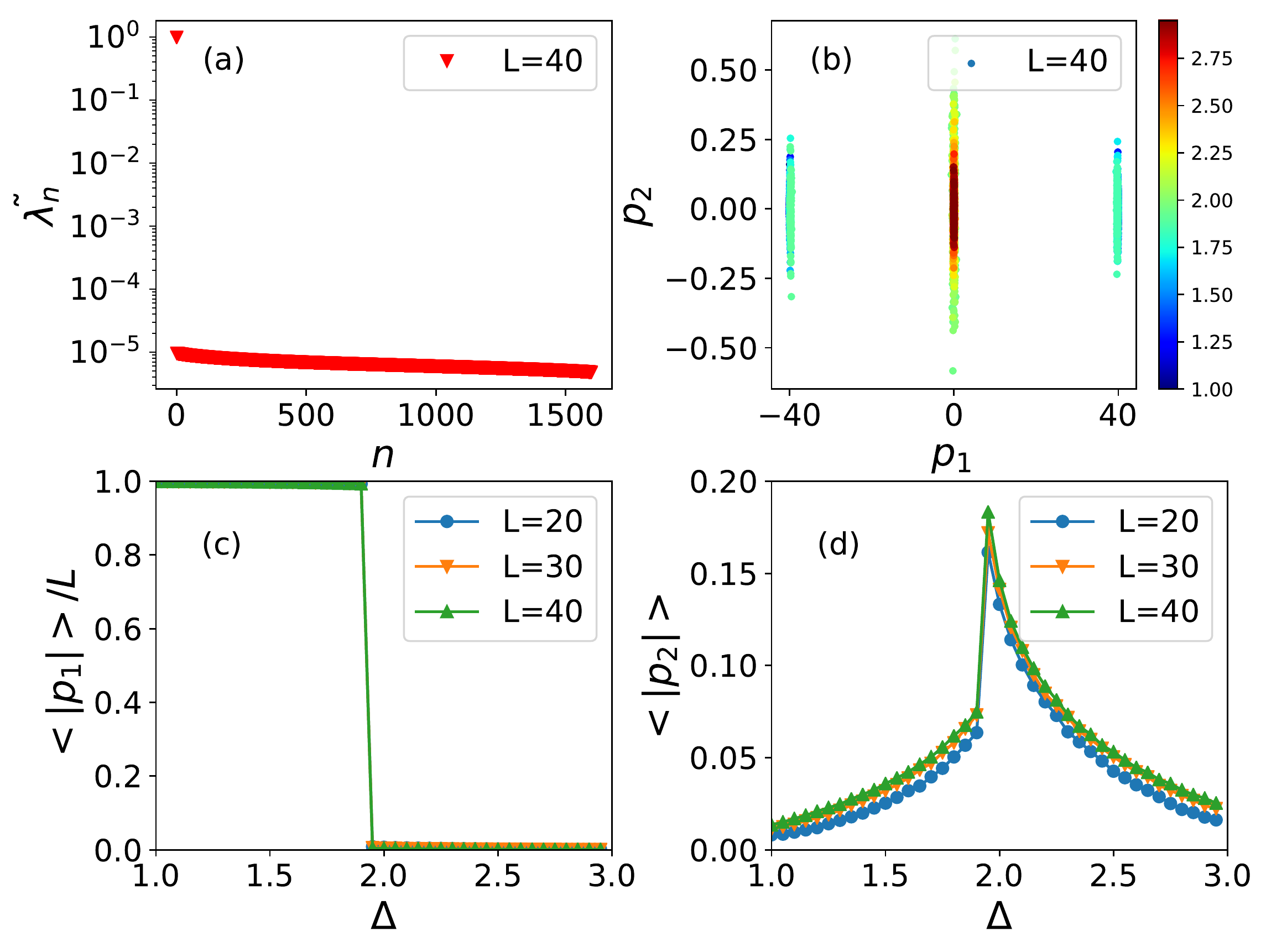}  
\caption{
PCA results for the BCM at fixed $J=1.0, T=0.4$.  (a)
Relative variances. (b) Projection of the
raw Blume-Capel configurations onto the plane of two leading principal
components. The color bar on the right
indicates $\Delta$ in units of $J$. (c) The normalized quantified
first leading component as a function of $\Delta$. (d) The quantified
second leading component as a function of $\Delta$.
Peaks provide estimates of transition value $\Delta_c$. 
\label{fig:Blume_Capel_T04}
}
\end{figure}

Following the same procedure, we feed a set of raw Blume-Capel 
spin configurations
$\{\mathbf{S}_i\}$ into the PCA to generate the results shown in
Figs.~\ref{fig:Blume_Capel_T10}, \ref{fig:Blume_Capel_T04}.
For both plots, we choose $t=40$ values of $\Delta$ and $n=10,000$
uncorrelated spin configurations for each $\Delta$. 
As noted earlier, 
the phase diagram of the BCM differs from that of
the Ising model in one important respect.  It
possesses both first and second order
transitions, separated by a tricritical point at 
$(T/J,\Delta/J)=(0.609(4),1.965(5))$.
In Fig.~\ref{fig:Blume_Capel_T10}, the PCA was
supplied with a set of spin configurations in a
sweep of $\Delta$ in the range
$0.5 < \Delta < 2.5$ at fixed $T=1.0$, thus crossing
the phase boundary in a second order transition.
Similarly, in Fig.~\ref{fig:Blume_Capel_T04} the sweep 
covers the range $1.0 < \Delta < 3.0$ at fixed $T=0.4$, and
crosses the phase boundary in a first order transition.

The panels of Fig.~\ref{fig:Blume_Capel_T10} are very similar
to that of the Ising model, 
Fig.~\ref{fig:Regular_Ising_Model}.
Using the peaks returned from
Fig.~\ref{fig:Blume_Capel_T10} (d) and finite size scaling, we can
locate $\Delta_c \approx 1.70 \pm 0.01$ in the thermodynamic limit. 
This is close to the value $\Delta_c \approx 1.63$ at 
$T=1.0$ reported in \cite{Pittman08}. 
Of course, to
locate the critical point more accurately, one could refine the
$\Delta$ range and add more samples at each temperature, much
as improvements in critical values are obtained with conventional analysis.

At fixed $T=0.4$, the BCM undergoes a first order
transition at $\Delta_c \approx 1.996$\cite{Kwak}.
The corresponding PCA
data are shown in Fig.~\ref{fig:Blume_Capel_T04}. These first order phase
transition results can be easily distinguished from those
at the second order transition, which were seen in
Figs.~\ref{fig:Regular_Ising_Model} (b), \ref{fig:Blume_Capel_T10} (b)
to spread out uniformly from the origin.  In stark contrast,
the scatter
points in Fig.~\ref{fig:Blume_Capel_T04} (b) spread out mostly in the
second principal component, and there are few intermediate points
between two phases. The latter observation is of course
expected from a first order transition. 
We see that, PCA clearly behaves differently across first
and second order transitions.

%%%%%%%%%%%%%%%%%%%%%%%%%%%%%%%%%%%%%%%%%%%%%%%%%%%%%%%%%%%%%%%%%%%%
\subsection{Monte Carlo and PCA results of the BSI model}
%%%%%%%%%%%%%%%%%%%%%%%%%%%%%%%%%%%%%%%%%%%%%%%%%%%%%%%%%%%%%%%%%%%%

\begin{figure}[!h]
\includegraphics[width=0.98\columnwidth]{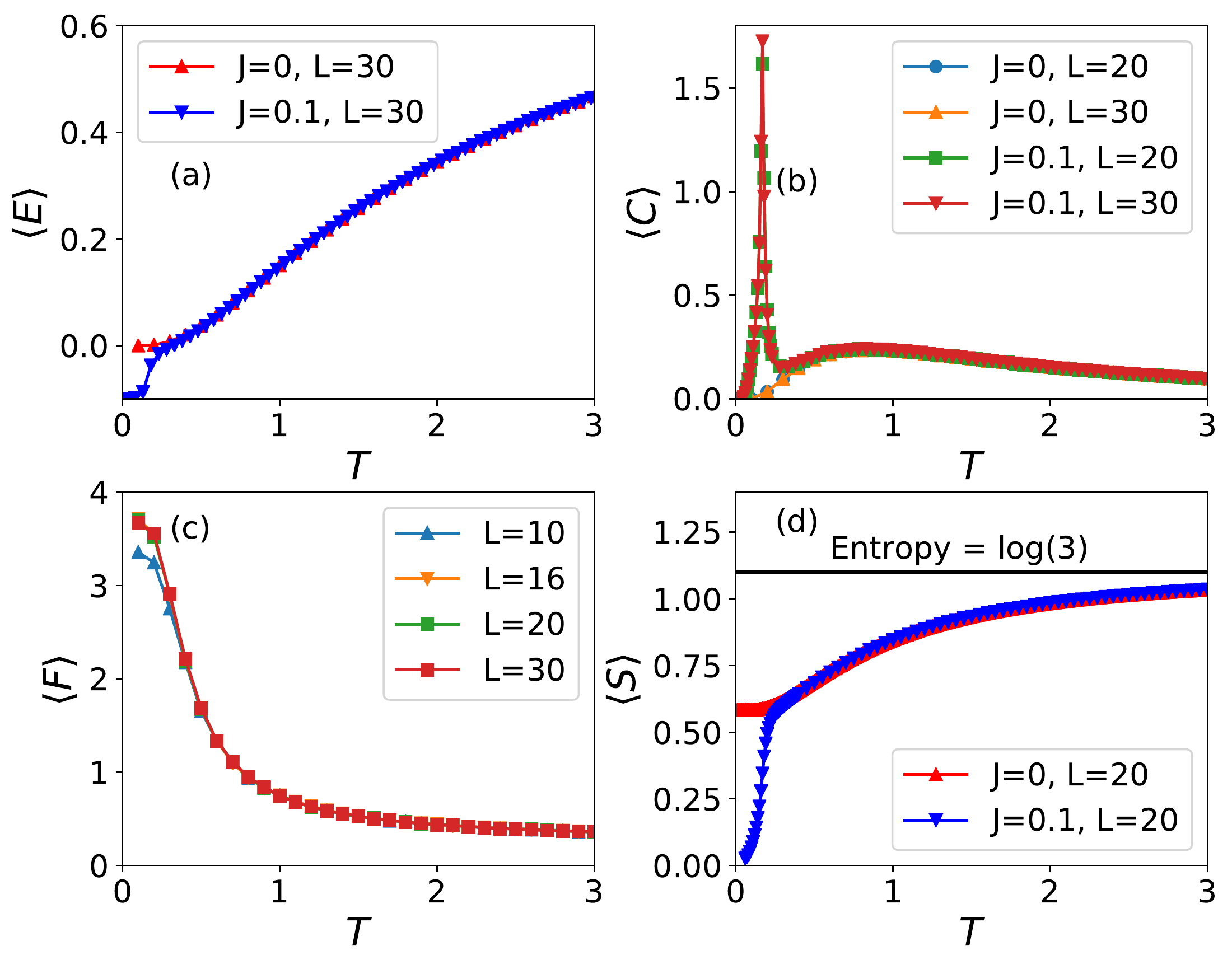}  
\caption{
(a) Energy as a function of temperature with (Blue) or without (red) $J$
term. (b) Specific heat as a function of temperature with (green, brown)
or without (blue, yellow) $J$ term. Finite size effect is verified to be
negligible. Sharp peaks agree well with the Binder ratio crossing in
Fig.~\ref{fig:Modified_Ising_Model_3} (c). (c) Structure factor as a function of temperature with different system sizes. $J$ is fixed to $0$. (d) Entropy as a function of
temperature with (blue) or without (red) the $J$ term. Ground-state
entropy for the red curve is $0.584$. 
\label{fig:Modified_Ising_Model_2}
}
\end{figure}

To explore thermodynamic properties of the BSI model beyond the
preliminary results presented in Fig.~\ref{fig:Modified_Ising_Model_1}
we first study its energy, specific heat and the `checkerboard'
structure factor, which are shown in
Fig.~\ref{fig:Modified_Ising_Model_2} (a), (b) and (c). Their
definitions are as follows:

\begin{align}
\left\langle E \right\rangle &= (1/N')\sum_{\sigma_i}(E_{\sigma_i}/N) 
\nonumber \\
\left\langle C \right\rangle &= 
N\frac{\left\langle E^2 \right\rangle - \left\langle E \right\rangle^2}{T^2}
\nonumber \\
\left\langle F \right\rangle &= 
(1/N)\sum_{\mathbf{i}, \mathbf{j}}e^{i\mathbf{Q}(\mathbf{i}-\mathbf{j})}
(\left\langle S^2_{\mathbf{i}} S^2_{\mathbf{j}} \right\rangle 
- \left\langle S^2_{\mathbf{i}} \right\rangle \left\langle S^2_{\mathbf{j}} 
\right\rangle)
\label{eq:BSI_eq_E_C_F}
\end{align}
where $\sigma_i$ is the spin configuration from the Monte Carlo
simulation, $E_{\sigma_i}$ is the total energy of the spin configuration
$\sigma_i$, $N'$ is the total number of spin configurations used for calculation and $\mathbf{Q}$ is the ordering 
wavevector $(\pi,\pi)$

With only the biquadratic term, the specific heat of the BSI model has a
peak at $T\approx 0.8$, however, the intensity of the peak doesn't
increase with system size, indicative of only short-range order.  The
structure factor is shown in Fig.~\ref{fig:Modified_Ising_Model_2} (c).
It shows a sharp increase around the same temperature as the specific
heat peak.  However, once again, the peak doesn't grow with system size,
indicating an absence of a phase transition for the $J=0$ BSI model.  If
we turn on the next nearest neighbor bilinear interaction ($J=0.1$), we
can see a clear sharp peak in the specific heat at $T\approx 0.17$ which
grows with system size, indicating a true phase transition.  This peak
is very sharp because the strength of the bilinear exchange interaction
$J=0.1$ is small compared with the leading energy scale $K=1$. 

For frustrated systems, an important quantity is the entropy of the system.
In Fig.~\ref{fig:Modified_Ising_Model_2} (d), we show the
entropy, calculated from:
\begin{small}
\begin{align}
S(T_b)-S(T_a) &=\int^{T_b}_{T_a} \frac{ C(T')}{T'}dT'\nonumber \\
&=\frac{E(T_b)}{T_b} - 
\frac{E(T_a)}{T_a} +
\int^{T_b}_{T_a} \frac{E(T')}{T'^2}dT'
\label{eq:BSI_eq_S}
\end{align}
\end{small}

Replacing $E$ with the Monte Carlo average $\left\langle E
\right\rangle$ and approximating $S(T_b)$ with
$S(+\infty)-c/T^2_b$, where $c$ is a fitting constant, we can
estimate the low temperature entropy with:
\begin{align}
\left\langle S(T) \right\rangle &= \left\langle S(+\infty) \right\rangle -\frac{c}{T^2_b} \nonumber \\
&-\frac{\left\langle E(T_b) \right\rangle}{T_b} + 
\frac{\left\langle E(T_a) \right\rangle}{T_a} - \int^{T_b}_{T_a} \frac{\left\langle E(T') \right\rangle }{T'^2}dT'
\label{eq:BSI_eq_S2}
\end{align}

With only the biquadratic term, the system has a non-zero ground-state
entropy of approximately $0.584$, which clearly exceeds $\ln{3}/2\approx
0.5493$ obtained from fixing one sublattice to have $S_i=0$ and allowing
spins to take arbitrary values on the other sublattice.  With a non-zero
$J=0.1$, below $T_c$ the entropy quickly drops to zero. For the Ising
antiferromagnet on a triangular lattice, we have checked that we obtain
the correct ground state entropy\cite{Wannier, Stephenson1,
Stephenson2}, with the value $0.323$, as reported in
Ref.~\onlinecite{Singh}.

\begin{figure}[!h]
\includegraphics[width=0.98\columnwidth]{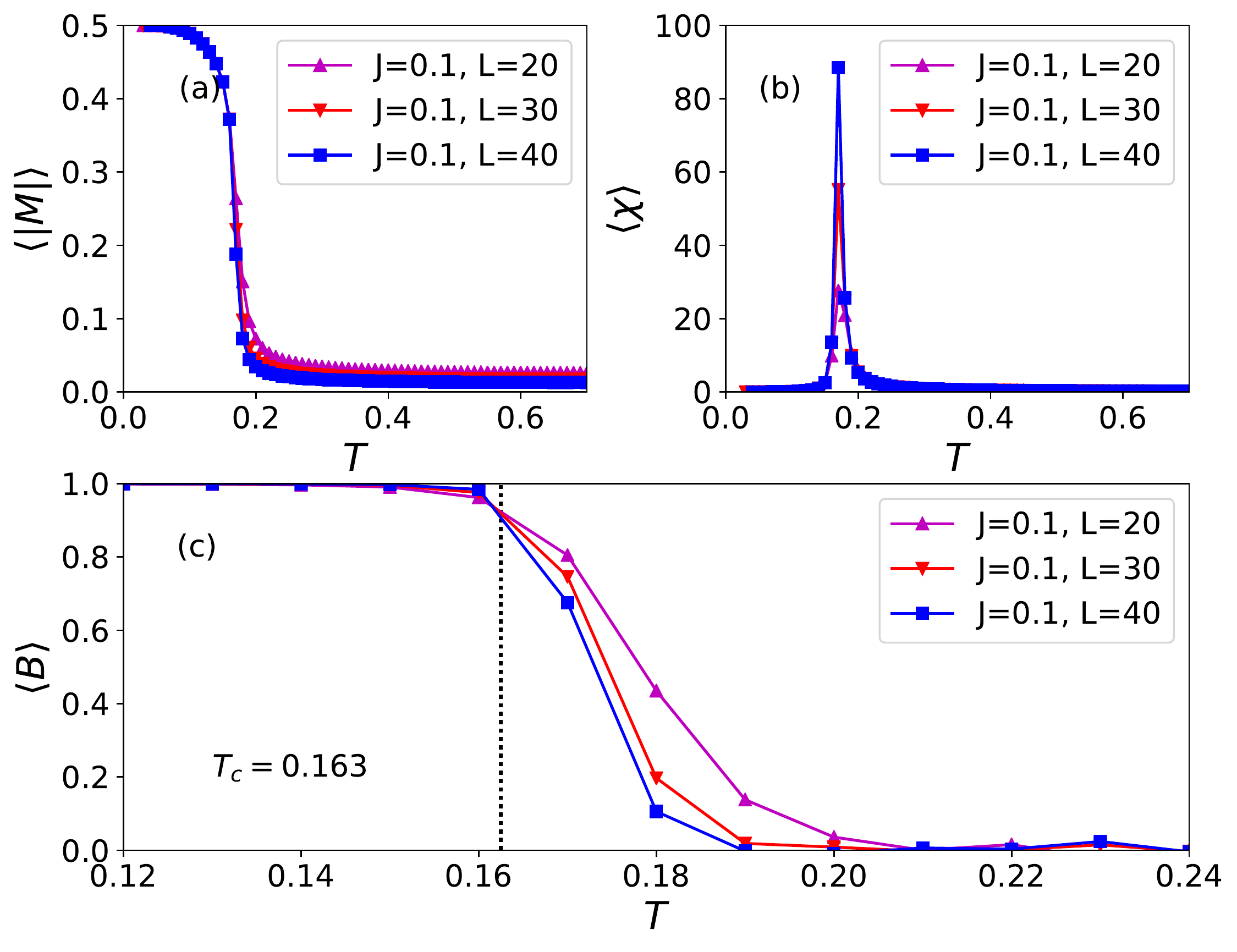}  
\caption{
Monte Carlo results for the BSI model with fixed $K=1.0$, $J=0.1$. (a)
Absolute magnetization as a function of temperature. (b) Susceptibility
as a function of temperature. Sharp peaks agree well with the Binder
ratio crossing. (c) Binder ratio as a function of temperature with
varied system sizes. The critical temperature is $T_c=0.163$. 
\label{fig:Modified_Ising_Model_3}
}
\end{figure}

To locate the critical temperature accurately in the latter case, we
calculate the absolute magnetization $\left\langle |M| \right\rangle$,
the susceptibility $\left\langle \chi \right\rangle$ and the Binder
ratio\cite{Binder} $\left\langle B \right\rangle$, which are shown in
Fig.~\ref{fig:Modified_Ising_Model_3} and defined as:
\begin{align}
\left\langle |M| \right\rangle &= (1/N')\sum_{\sigma_i}(|\sum_j s_j|/N) 
\nonumber \\
\left\langle \chi \right\rangle &= 
N\frac{\left\langle M^2 \right\rangle - \left\langle |M| \right\rangle^2}{T}
\nonumber \\
\left\langle B \right\rangle &= 
\frac{1}{2}\Big(3-\frac{\left\langle M^4 \right\rangle}
{\left\langle M^2 \right\rangle}\Big)
\label{eq:BSI_eq_M_chi_B}
\end{align}
where $N'$ is the total number of spin configurations used for the
calculation. Similar to $\left\langle C \right\rangle$, both
$\left\langle |M| \right\rangle$ and $\left\langle \chi \right\rangle$
have small finite size effects and their transition signals agree
well with the peak of $\left\langle C \right\rangle$. In
Fig.~\ref{fig:Modified_Ising_Model_3} (c), the Binder ratio plot shows a crossing at $T_c=0.163$, which gives the critical point for the $J=0.1$ BSI model.

\begin{figure}[!h]
\includegraphics[width=0.98\columnwidth]{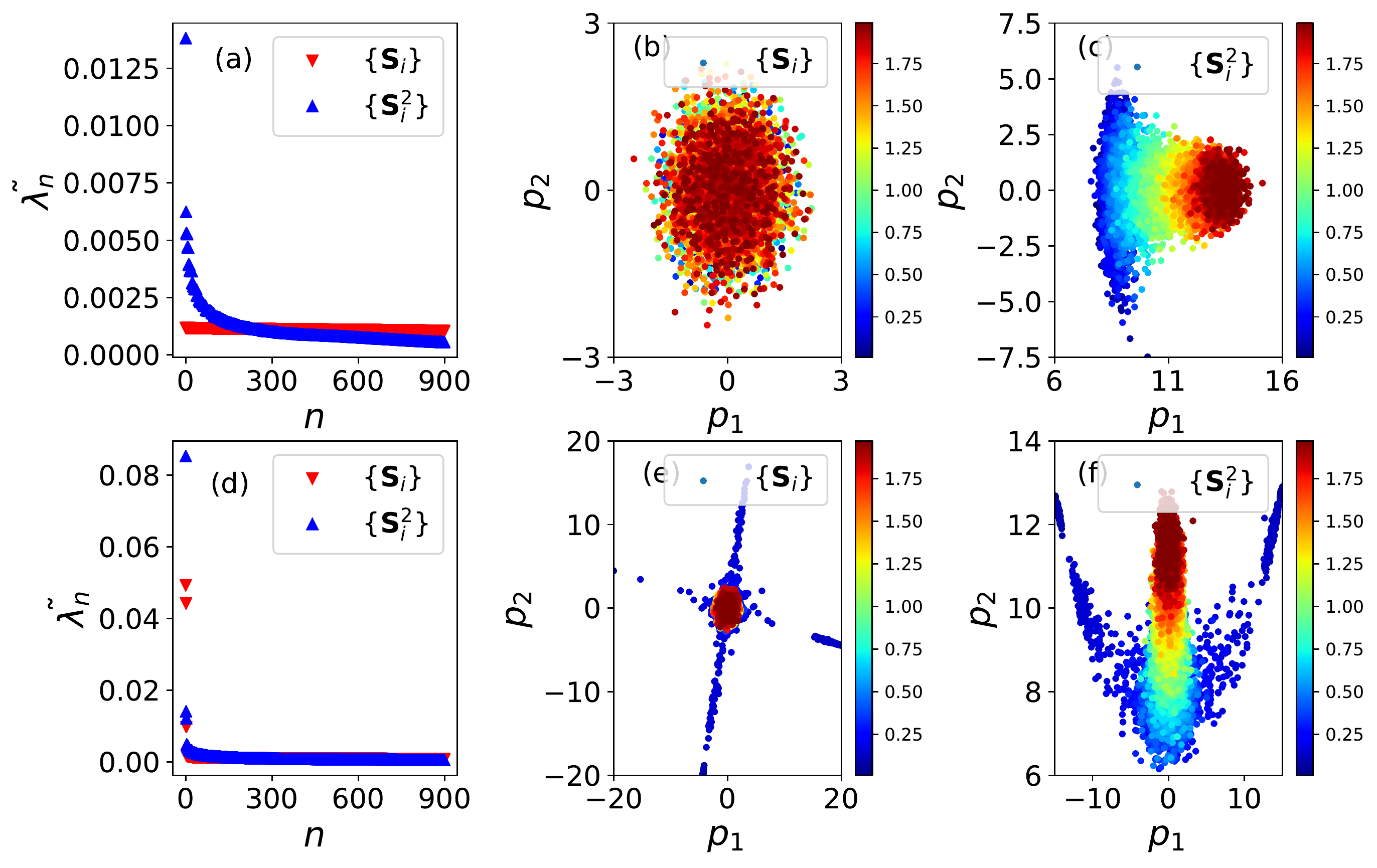}  
\caption{
PCA results for the BSI model at lattice size $L=30$. Color bars of
scatter plots indicate the temperature of the samples in the unit of
$K$. \underline{Top row:} The BSI model with fixed $K=1.0$, $J=0$.  (a) Relative variances obtained from the raw BSI configurations
$\{\mathbf{S}_i\}$ (red) and the squared spin configurations
$\{\mathbf{S}^2_i\}$ (blue). (b) Projection of the raw BSI
configurations $\{\mathbf{S}_i\}$ onto the plane of leading two
principal components. (c)Projection of the squared spin configurations
$\{\mathbf{S}^2_i\}$ onto the plane of leading two principal components. 
\underline{Bottom row:} The BSI model with fixed $K=1.0$, $J=0.1$.  (d),
(e), (f) follow same definitions as (a), (b), (c).
\label{fig:Modified_Ising_Model_4}
}
\end{figure}

Next, we apply the PCA method to the BSI model and generate results as
shown in Fig.~\ref{fig:Modified_Ising_Model_4}. In the top row of
Fig.~\ref{fig:Modified_Ising_Model_4}, we display PCA results for the
$J=0$ case. We first feed raw spin configurations $\{\mathbf{S}_i\}$
into PCA, however, no dominant principal components are observed (see
red symbols).  Projecting raw spin configurations onto the leading
principal components plane, we observe a complete mixture of points of
all temperatures.  This is one of the failures of an automated machine
learning algorithm, which we will see again in the study of the XY
model. The physics of the problem relates to how the $S_i=\pm 1$ spin
variables arrange with respect to the $S_i=0$ variables, not with
respect to each other. But, PCA averages these $\pm 1$ variables out to
zero and fails to recognize any emergent order.

However, if we feed the squared spin configurations $\{\mathbf{S}^2_i\}$
into PCA, we can get dominant principal components (blue symbols).  If
we further project squared spin configurations into two leading
principal components, as shown in Fig.~\ref{fig:Modified_Ising_Model_4}
(c), we are able to distinguish the low temperature phase with the high
temperature phase. Actually, the first leading principal component is
just the local `charge' order parameter, where we take $s_i=+1$ and
$s_i=-1$ as `charge' one and $s_i=0$ as `charge' zero. The charge
density changes from $2/3$ at high temperatures to below $0.5$ at low
temperatures. PCA analysis shows that this as a greater discriminant of
different temperatures than the true emergent charge order in the
checkerboard pattern. The checkerboard order forms only the second
principal component. Although it has a larger relative change with
temperature, in magnitude it is smaller than the extensive change in the
total number of charges present.

The scatter points in Fig.~\ref{fig:Modified_Ising_Model_4} (c) are
different from the case of first order phase transition in
Fig.~\ref{fig:Blume_Capel_T04} (b) or the second order phase transition
in Fig.~\ref{fig:Blume_Capel_T10} (b).  The most evident difference is
the absence of symmetry breaking, which in
Fig.~\ref{fig:Blume_Capel_T04} (b) and Fig.~\ref{fig:Blume_Capel_T10}
(b) shows up by the bifurcation of scatter points for the low
temperature data at left and right ends. This is clear signature for a
gradual crossover to a low temperature behavior and an absence of a
phase transition.

\begin{figure}[!h]
\includegraphics[width=0.98\columnwidth]{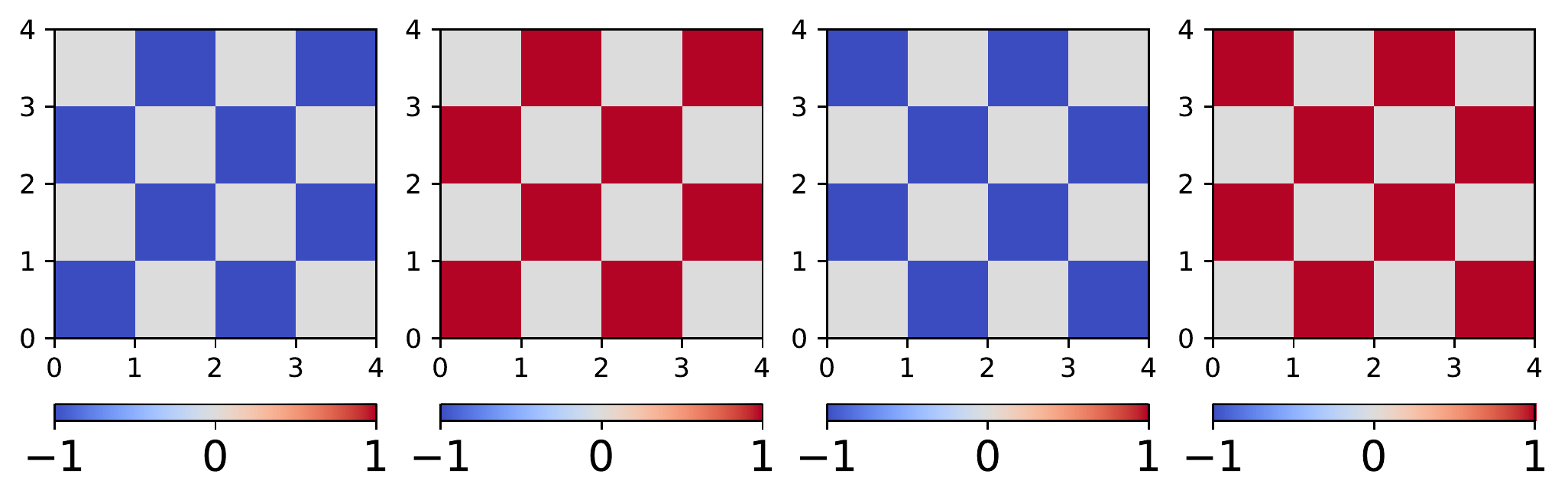}  
\caption{
Four distinct ground-states for the BSI model, with fixed $K=1$,
$J=0.1$, $N=40\times 40$. For clarity, we only display a $4\times 4$ block
for each ground-state.
\label{fig:Modified_Ising_Model_5}
}
\end{figure}

Turning on the bilinear exchange interaction, by feeding either the raw
spin configurations $\{\mathbf{S}_i\}$ or the squared spin
configurations $\{\mathbf{S}^2_i\}$ into PCA, we observe dominant
principal components, as shown in
Fig.~\ref{fig:Modified_Ising_Model_4} (d). There is only one
dominant relative variance returned from $\{\mathbf{S}^2_i\}$ but two equally weighted dominant relative variances returned from $\{\mathbf{S}_i\}$. 
In Fig.~\ref{fig:Modified_Ising_Model_4} (e), projecting raw spin
configurations onto the plane of two leading principal components
$p_{i1}$ and $p_{i2}$, we see four distinct branches extending from the
center, and interestingly, all of them correspond to temperatures lower
than $T_c$. In this model, there exist four distinct ground-states, which are
shown in Fig.~\ref{fig:Modified_Ising_Model_5}, and each
distinct ground-state is associated with a branch in
Fig.~\ref{fig:Modified_Ising_Model_4} (e). 
Thus we see that the number of
dominant relative variances $N_{v}$ are related to the number of
ground-states $N_{g}$ by $N_{g}=2^{N_{v}}$.

On the other hand, by feeding $\{\mathbf{S}^2_i\}$ into PCA and
projecting data onto two leading principal components, we generate
Fig.~\ref{fig:Modified_Ising_Model_4} (f). The original four branches
become two, because replacing $s_i$ with $s_i^2$, will only leave us
with two distinct ground-states corresponding to the charge order. In
this case, it is evident that $s_i^2$ can not capture the full phase
behavior of the model.

%%%%%%%%%%%%%%%%%%%%%%%%%%%%%%%%%%%%%%%%%%%%%%%%%%%%%%%%%%%%%%%%%%%%
\subsection{PCA results of the TLIM }
%%%%%%%%%%%%%%%%%%%%%%%%%%%%%%%%%%%%%%%%%%%%%%%%%%%%%%%%%%%%%%%%%%%%

\begin{figure}[!h]
\includegraphics[width=0.98\columnwidth]{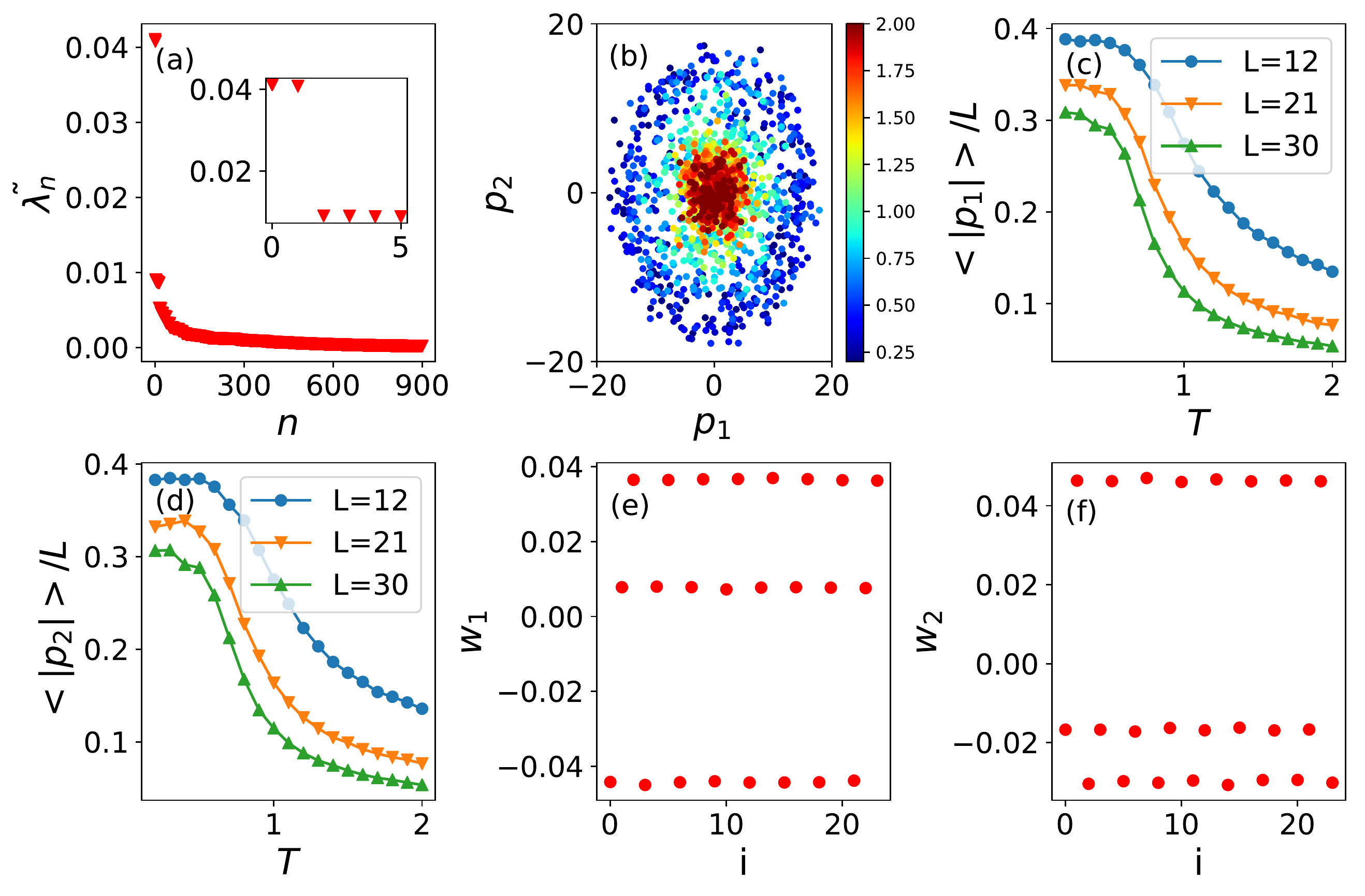}  
\caption{
\underline{Top row:} PCA results for the antiferromagnetic (fully
frustrated) TLIM, with fixed $J=-1$. (a) Relative variances. $L=30$. (b)
Projection onto the plane of leading two principal components. The color
bar indicates the temperature in the unit of $|J|$. $L=30$. (c) The
normalized quantified first leading component as a function of
temperature. (d) The normalized quantified second leading component as a
function of temperature. (e) The weight vector corresponding to the
first leading component.  (f) The weight vector corresponding to the
second leading component.  See text for interpretation.
\label{fig:crossover_models}
}
\end{figure}

We now turn to the PCA analysis of the fully frustrated
antiferromagnetic Triangular-Lattice Ising Model (TLIM).  This is a
particularly interesting case as there is no long-range order in the
model all the way down to $T=0$. Instead, power-law spin correlations
develop as temperature is lowered \cite{Stephenson1, Stephenson2}.  We
feed raw spin configurations $\{\mathbf{S}_i\}$ into PCA. We observe two
equally weighted relative variances, as shown in
Fig.~\ref{fig:crossover_models} (a).  If we project them into leading
two principal components, we see a separation of high temperature
scatter points, forming a central blob, from the low temperature scatter
points, forming the outside periphery of the circle, as shown in
Fig.~\ref{fig:crossover_models} (b). Figures (c) and (d) show the growth
of the projected principal components as the temperature is lowered.
Figures (e) and (f) show the spatial patterns in the projected
components.

It is interesting to observe that the two principal components
correspond to patterns of ordering in which the three sub-lattices of
the triangular-lattice form roughly an (m,0,-m) and an (m,-m/2,-m/2)
pattern respectively.  PCA analysis shows an emergent continuous XY type
symmetry between these patterns. It is known \cite{moessner,grest} that
these are the two incipient orders of the triangular antiferromagnet
when either a weak transverse-field is added to the system or if
triangular-Ising layers are stacked on top of each other with a weak
interlayer coupling. These two patterns are nearly degenerate. Within
the Ginzburg-Landau approach, the degeneracy is only lifted by an
irrelevent 6th order anisotropy term. It is remarkable that machine
learning can automatically point to such incipient order in the Monte
Carlo data. Independently recognizing such incipient patterns may be one
of the strengths of machine learning.

%%%%%%%%%%%%%%%%%%%%%%%%%%%%%%%%%%%%%%%%%%%%%%%%%%%%%%%%%%%%%%%%%%%%
\subsection{PCA results of the XY model}
%%%%%%%%%%%%%%%%%%%%%%%%%%%%%%%%%%%%%%%%%%%%%%%%%%%%%%%%%%%%%%%%%%%%

\begin{figure}[!h]
\includegraphics[width=0.98\columnwidth]{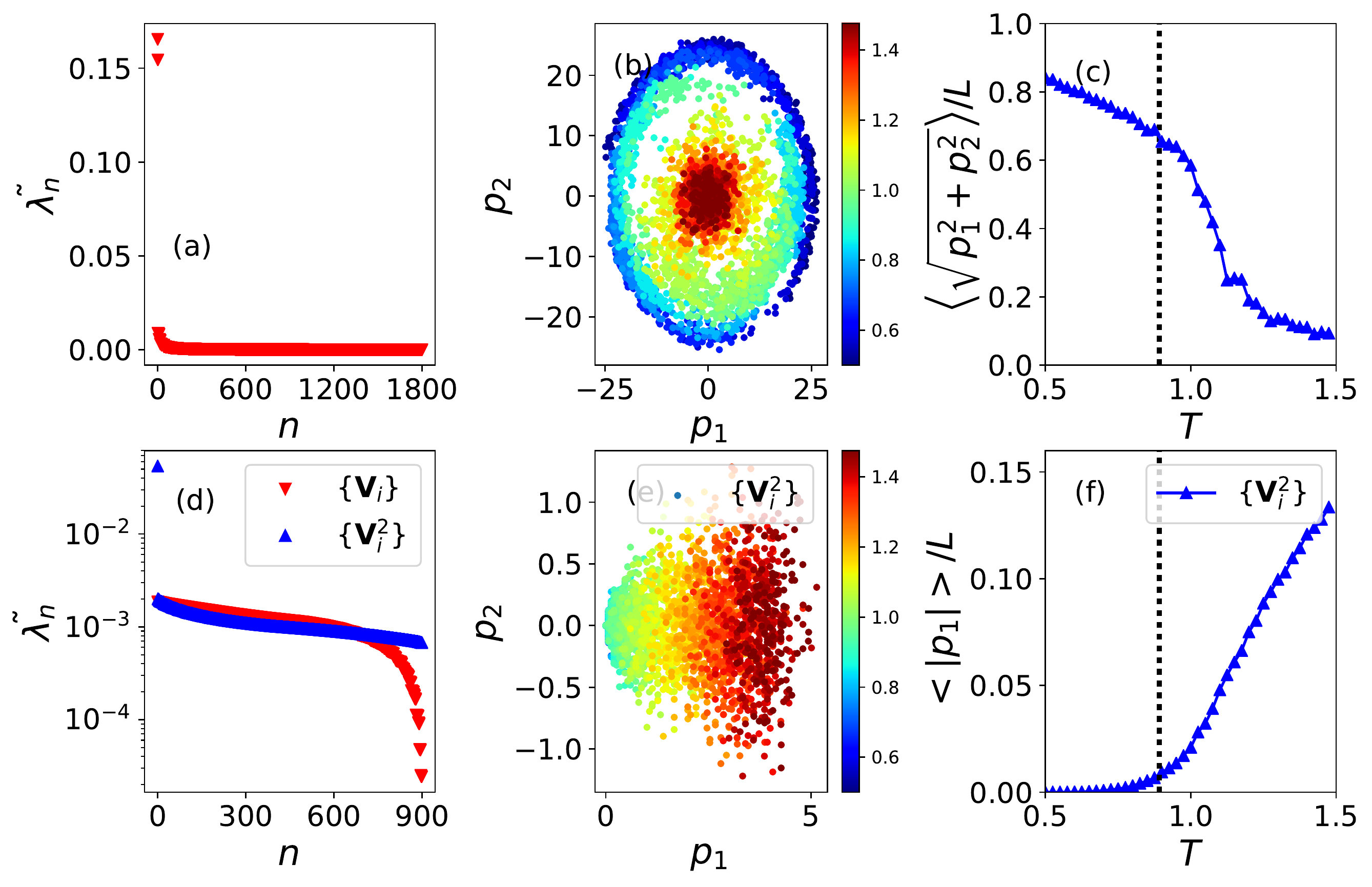}  
\caption{
PCA results for the XY model with fixed $J=1.0$ at $L=30$.
\underline{Top row:} Feed angle configurations $\{\cos(\theta_i),
\sin(\theta_i)\}$ into PCA.  (a) Relative variances. (b) Projection onto
the plane of leading two principal components. The color bar indicates
the temperature in the unit of $J$. Each temperature has $100$ scatter
points. (c) $\left\langle\sqrt{p^2_1+p^2_2}\right\rangle /L$ versus
temperature. The dash line marks the true $T_c$. \underline{Bottom row:}
(d) Relative variances. Red (Blue) triangles, with the label
$\{\mathbf{V}_i\}$ ($\{\mathbf{V}^2_i\}$), indicate results by feeding
the raw vorticity (squared vorticity) configurations into PCA. (e)
Projection of the squared vorticity configurations onto leading two
principal components. The color bar indicates temperature and each
temperature has $100$ scatter points. (f)
$\left\langle|p_1|\right\rangle/L$ versus temperature. The dash line
marks the true $T_c$.
\label{fig:XY_model}
}
\end{figure}

We now focus on the PCA analysis of the 2D XY model by feeding various
sorts of real space snapshots, the spin directions themselves and
measures of local vorticity, into PCA. To express spin directions, we
define an angle configuration as $[\cos(\theta_1), \sin(\theta_1),\dots,
\cos(\theta_N), \sin(\theta_N)]$, and a collection of angle
configurations simply as $\{\cos(\theta_i), \sin(\theta_i)\}$. By
providing these projections of the XY
spins along x axis and y axis to PCA, we observe two equally weighted
relative variances in Fig.~\ref{fig:XY_model} (a) and concentric circles
in Fig.~\ref{fig:XY_model} (b), which essentially recovers the
rotational symmetry of XY spins in the spin space. In
Fig.~\ref{fig:XY_model} (b), we can see that in the high temperature phase,
scatter points aggregate in the center, and are separated from the low
temperature scatter points, which spread out in the periphery of the
circle.  In figure (c) we show a plot of $\left\langle
\sqrt{p^2_1+p^2_2} \right\rangle /L$, which is the order parameter for
the system.  Also shown in figure (c) by a dashed vertical line is the
known KT transition temperature $T_c=0.892$\cite{Olsson}.  The data show
an abrupt change in the temperature dependence around $T_c$.
Differentiating power-law correlations from long-range order is possible
in PCA but will require careful finite-size study of the projected
principal components, which will again take us back to conventional
analysis. 

The KT transition is known to be driven by unbinding of
vortex-antivortex pairs. Since, PCA is a linear transformation method, it
can not deduce vortex-related information from the spin configurations themselves.
However, we can first preprocess raw spin
configurations to generate local vortex and anti-vortex configurations
by the definition:
\begin{align}
\oint_C \Delta \theta dl = 2\pi k \hskip0.50in 
k=\pm 1, \pm 2, \pm3 \dots
\label{eq:XY_vorticity}
\end{align}
where $k$ is the winding number and the integral is done for each square
plaquette. Explicitly, each spin difference along a plaquette bond is rescaled into
the range $(-\pi, \pi]$.  The spin differences are then summed over the plaquette
to give the `integral' of Eq.~\ref{eq:XY_vorticity}.  The winding number
$k$ is non-zero only when the integral is a multiple of $2\pi$. Thus
each square plaquette is expressed using its winding number, which is
either $+1$, indicating a vortex, $0$, indicating no vorticity, or $-1$,
indicating an anti-vortex.  However, when we feed the raw vorticity
configurations into PCA, we do not observe any dominant principal
components, as shown in Fig.~\ref{fig:XY_model} (d) red triangles. This
again shows the limitation of the PCA method.  If equally weighted
positive and negative vorticities exist in one configuration, they will
cancel out each other, and PCA fails to capture their proliferation with
temperature.  This is analogous to the failure of PCA in capturing the
`charge' density in the BSI model when feeding $\{\mathbf{S}_i\}$.

On the other hand, if we feed absolute vorticity configurations (defined
to be +1 if there is a vortex or antivortex present and 0 otherwise)
into PCA, a clear dominant principal component is observed. See blue
triangles in Fig.~\ref{fig:XY_model} (d). The scatter points of the
leading two principal components are shown in figure (e), which appears
more like a crossover from previous examples than a true phase
transition. We believe, seeing the vortex-antivortex unbinding
transition in the 2D XY model may be difficult in PCA or any machine
learning technique.  The change of projected principal component with
temperature is shown in Fig.~\ref{fig:XY_model} (f) and it shows the
proliferation of vortices with increase in temperature. The dashed
vertical line again shows the known transition temperature and is
consistent with the temperature of onset for proliferation of vortices.

A clear message from these studies is that PCA with spatial
configurations of local variables is primarily a study of relevant
patterns, their symmetry, and symmetry-breaking in the model.  The
discrete symmetry breaking shows up in the 2 or 4-fold patterns of
separation of principal component data. Continuous rotational symmetry
in the spin or order parameter space is reflected in the circular
pattern of scatter points.

%%%%%%%%%%%%%%%%%%%%%%%%%%%%%%%%%%%%%%%%%%%%%%%%%%%%%%%%%%%%%%%%%%%%
\subsection{Autoencoder results of the Ising model}
%%%%%%%%%%%%%%%%%%%%%%%%%%%%%%%%%%%%%%%%%%%%%%%%%%%%%%%%%%%%%%%%%%%%

\begin{figure}[!h]
\includegraphics[width=0.98\columnwidth]{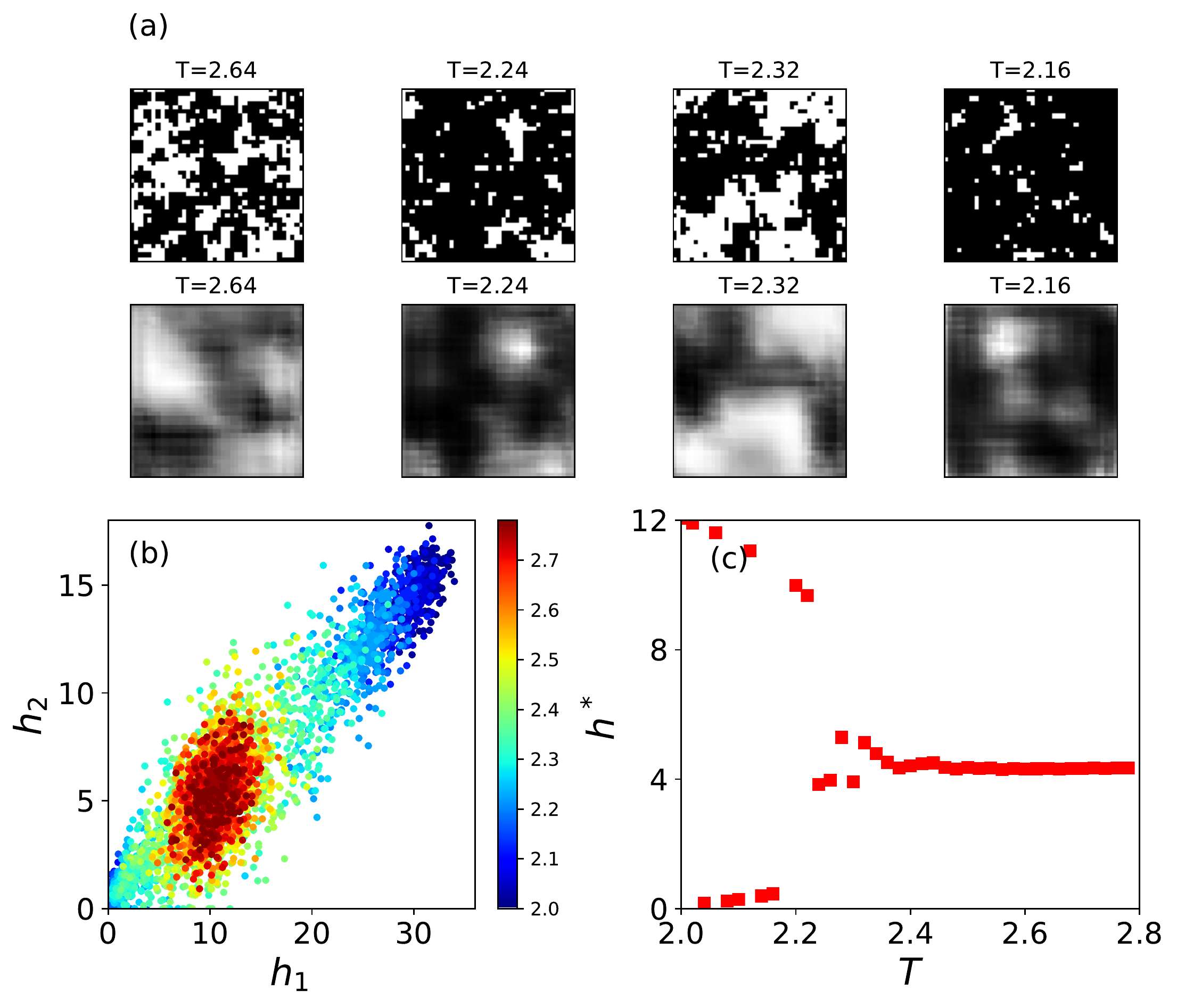}  
\caption{
Ising model autoencoder results. $J=1.0$ is fixed and the original lattice size is $N=40\times 40$. (a) Spin
configuration (above) and reconstructions (below) at the indicated
temperatures using $200$ hidden neurons. (b) Encoding of the raw Ising
configurations $\{\mathbf{S}_i\}$ onto the plane of the two hidden
neuron activations ($h_1$, $h_2$). Color bar indicates the temperature of the sample in units of
$J$. Each temperature has $100$ scatter points. (c) Encoding of the raw
Ising configurations $\{\mathbf{S}_i\}$ using a single hidden neuron activation ($h^*$) as
a function of temperature.  Each data point is averaged over $n=10000$
samples of the same temperature.
\label{fig:autoencoding_3}
}
\end{figure}

The autoencoder method and architecture are explained in Sec.~IIIB and
the Appendix.  In Fig.~\ref{fig:autoencoding_3} (a), we display
reconstruction results using $200$ hidden neurons. The top row is the
raw Ising configurations at several temperatures and the bottom row is
the reconstructed Ising configurations. Although the reconstructed
configurations lose detailed spin structures, the network succeeds to
capture essential informations of the input data, such as the domain
area, the domain position and also the overall magnetization.  

Next, we limit the autoencoder network to only two hidden neurons and
results have been shown in Fig.~\ref{fig:autoencoding_3} (b).
Interestingly, we obtain similar results with the PCA method, as shown
in Fig.~\ref{fig:Regular_Ising_Model}, and the ferromagnetic phase is
reasonably distinguished from the high temperature phase. 
%\textcolor{red}{The figure caption says `head neurons'.  Is that
%supposed to be `hidden neurons'?  Also, it is not really clear what
%$h_1$ and $h_2$ are.  They are the state of the two hidden neurons?
%Similarly in Fig.~13c is $h_*$ the value of the single hidden
%neuron?  Do we understand why it is $h_*=0$ or $h_*=12$?}

Furthermore, we limit the autoencoder network with a single hidden
neuron and we still try to reconstruct original inputs, with results
revealed in Fig.~\ref{fig:autoencoding_3} (c). From the plot, we can
clearly observe that the transition temperature is around $2.3$, above
which, the autoencoder network learns a trivial mapping function by
projecting all high temperature configurations into an almost constant.
However, when the temperature is below the transition temperature, the
trained network is able to capture two distinct branches, which actually
corresponds to two distinct ground-states of the Ising model.  Thus,
even with a single hidden neuron, the autoencoder network can be trained
to learn key phase transition informations and help locate the
critical point.

%%%%%%%%%%%%%%%%%%%%%%%%%%%%%%%%%%%%%%%%%%%%%%%%%%%%%%%%%%%%%%%%%%%%%%%%%%
\section{Conclusions}
%%%%%%%%%%%%%%%%%%%%%%%%%%%%%%%%%%%%%%%%%%%%%%%%%%%%%%%%%%%%%%%%%%%%%%%%%%

In this work, we have employed unsupervised machine learning techniques,
such as PCA and autoencoder, to study phase transitions in the Ising,
Blume-Capel, BSI and XY models. One of our key goals was to examine
critically what these methods are capturing from the Monte Carlo data
and as a result where they succeed and where they fail. In the variety
of models we studied here, some had a clearly discernible order
parameter.  In others it was a more subtle one. Some had no finite
temperature phase transition just a gradual crossover from uncorrelated
high temperature behavior to strongly correlated low temperature
behavior, some had a power-law ordered phase, and others a long-range
ordered one. Some had a first order phase transition, others a second
order one. Some had continuous symmetry, others a discrete symmetry. The
number of ordered phases varied between models.  It is clear from our
study that even a straightforward implementation of PCA or the
autoencoder-network machine-learning methods can be useful tools for
studying phase transitions and can distinguish between many different
scenarios.

%We also show, through an application to the Ising model, that 
%an autoencoder network can similarly be trained to locate
%phase transitions.

%% The BSI model has an unexplored phase diagram, so we first
%% tackled it with conventional Monte Carlo methods.  These approaches show
%% that the BSI model displays a crossover when only the biquadratic
%% term in nonzero.  A true phase transition emerges when the
%% next-nearest neighbor interaction is turned on.   Machine learning
%% methods, which we then applied, gave  useful information
%% about this subtle situation where many ground states
%% are present, with true order arising from a degeneracy-lifting term.

%%An obvious question is whether these machine learning methods
%% can compete with more
%% conventional methods, for example the use of Binder crossings
%% to locate the critical temperature of the Ising model \cite{Wolff, Robert, Katzgraber, Binder}. The results
%% presented here certainly do not come close to the accuracy of these well-developed
%% classical approaches.
%% An open question is whether, with time, improvement in these techniques can make them competitive.

Much of our focus was on the Principal Component Analysis (PCA) method. It is
clear that the PCA method, when one feeds in spin 
configurations from Monte Carlo data over a range of parameters, is primarily
about recognizing spatial patterns of order and symmetry breaking. This
is what discriminates one temperature from another. When we have a
rather obvious order parameter, such as in the square-lattice Ising model,
the dominant principal component must correspond to this order parameter.
The subleading principal components are related to small-q behavior,
and hence capture the domain-walls and low energy fluctuations in the system. 
Careful analysis of temperature and size dependence of the projections
on to these principal components can be used to obtain the transition 
temperature and the critical exponents, much like in conventional analysis.

In a fully frustrated system, such as the antiferromagnetic
triangular-lattice Ising model (TLIM), there is no obvious
order-parameter, only subtle incipient ones.  PCA automatically brings
them out. This may be one of the real strengths of automated
machine-learning methods. The TLIM has power-law correlations at T=0.
Furthermore, it has been shown\cite{moessner, grest} that a stack of
triangular-lattice Ising models with arbitrarily weak interlayer
coupling, or TLIM in a weak transverse quantum-field is susceptible to
order in a 3-sublattice pattern in which the three sublattices have
(m,0,-m) or (m,-m/2,-m/2) magnetization. These two patterns of order are
nearly degenerate, leading to an emergent XY symmetry which is only
broken by a higher order anisotropy term. PCA picks out these as the
principal components and the emergent XY symmetry between them. To fully
understand these one would clearly need further analysis. But, PCA could
serve as a useful starting point to the analysis of Monte Carlo data.

PCA can distinguish between first and second order transitions and between
crossovers and phase transitions. The most clear distinction comes in the
way the projections on to the lowest few principal components spread out.
The degeneracy or number of PCA components with large comparable variances 
and the groups in which the scatter points
of data arrange themselves very clearly and nicely
illustrate the number of ordered phases and
the symmetries of the model.

%One key feature of methods like the PCA is that they make use of
%only a small subset of the configurations of the system generated
%by the Monte Carlo simulation.  Since the accuracy of measurements
%in monte carlo increases as the square root of the sample
%size, it is natural to suppose that this truncation of information
%will limit accuracy.  On the other hand, conventional measures 
%which rely on a specific order parameter like the magnetization
%also truncate the information 
%by compressing the knowledge provided by the complete spin configuration
%to a single number.  It is possible that machine learning methods,
%while making use of fewer configurations, explore the 
%correlations of that subset to a greater degree of subtlety,
%specifically through information provided by the subleading
%components.

One glaring weakness of PCA is the inability to recognize an order
in $s_i^2$ when data on $s_i$ are fed in. We saw that problem in 
the BSI model, where short-range order is associated with 
checkerboard arrangement of charge ($s_i^2$)
not the original spin ($s_i$) variables. We encountered the same problem
in studying vortex-antivortex pairs in the 2D XY model. The PCA
fails to discriminate between configurations where opposite sign
charges equal in number are changing substantially with temperature. 
Only when the squared variables are fed into the PCA algorithms it
recognizes charge order or proliferation of vortices. It suggests that
it may generally be useful to apply PCA with different powers of the 
local spin-variable
to get a more complete picture.

The 2D XY model is particularly challenging for any numerical study. It
would be too much to expect an automated machine-learning algorithm to learn 
about unbinding of vortex-antivortex pairs from the data on raw spin 
configurations. In the current form, the algorithms can not even recognize 
the existence of vortices. Similarly, while we believe the power-law and 
exponentially decaying phases can be distinguished by studying the size 
dependence of the projected principal components, accurately locating
the phase-transition between them would be difficult. Designing machine
learning algorithms that can deal with such a situation will be an
important achievement.

An interesting question is whether automated machine-learning approaches
bear any parallels with the study of entanglement entropies in quantum
systems \cite{entanglement, Sarma}.
A major theme of condensed matter physics over the last decade has been
the exploration of phases with unusual, or perhaps even non-existent,
local patterns of order.  The entanglement entropy\cite{Osborne,
Calabrese, Cincio} of quantum lattice models is touted as a quantity
that, in principle, catches them all -- conventional order, conformal
invariance, Goldstone modes, Fermi-surfaces and topological order--
although, in practice, it takes substantial effort to address each of
these individually. It is clear from our study that the simple PCA,
without any pre-processing of data, is more closely tied to conventional
order. One can imagine defining a `discriminant-entropy' in terms of
the relative variances of the PCA method as a general characteristic of
the complexity of the model. It remains to be seen if such a properly
defined quantity can be helpful in studies of statistical models in an
order-parameter independent manner. This may be an interesting avenue
for future research.

Finally, we note that because they directly analyze the real
space configurations of the degrees of freedom, machine learning
algorithms might make useful connections to experimental
situations where the order varies over the sample.
For example, in studies of cold atomic gases, the confining
potential leads to a variation of density and energy scales
as one moves outward from the cloud center.  As a consequence,
metallic, magnetic, and Mott insulating phases can coexist in 
different regions of a single sample.  By allowing the machine learning
algorithm to be exposed to different spatial windows on real space data obtained from quantum microscopy\cite{Haller_Hudson, Bakr_Gillen, Sherson_Weitenberg}, useful
information about this coexistence and domain boundaries may be generated.

\begin{acknowledgements}
The work of WH and RTS was supported by DE-NA0002908.
The work of RRPS is supported by US National Science Foundation DMR-130608.
\end{acknowledgements}

%%%%%%%%%%%%%%%%%%%%%%%%%%%%%%%%%%%%%%%%%%%%%%%%%%%%%%%%%%%%%%%%%%%

\appendix*
%%%%%%%%%%%%%%%%%%%%%%%%%%%%%%%%%%%%%%%%%%%%%%%%%%%%%%%%%%%%%%%%%%%%
\section{Autoencoder Architecture}
%%%%%%%%%%%%%%%%%%%%%%%%%%%%%%%%%%%%%%%%%%%%%%%%%%%%%%%%%%%%%%%%%%%%

\begin{figure}[!h]
\includegraphics[width=0.98\columnwidth]{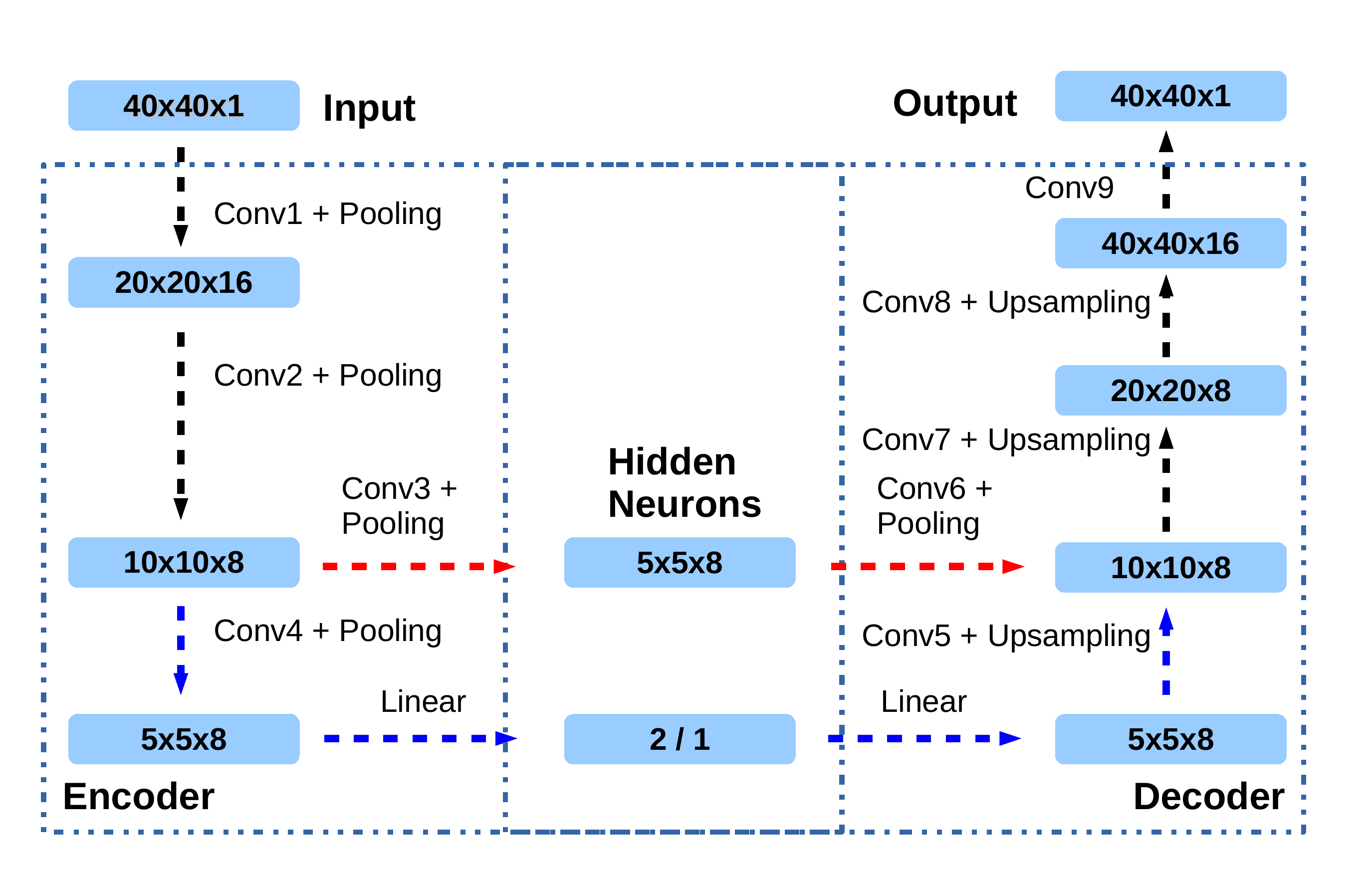}  
\caption{
Autoencoder architecture. To use $200$ hidden neurons, we choose to
follow the path from black to red to black, and to use $2$ or $1$ hidden
neuron, we choose to follow the path from black to blue to black. We use
raw spin configurations as the input data, with fixed size $N=40\times 40$.
\label{fig:autoencoding_2}
}
\end{figure}

In the preceding discussion, we have demonstrated detailed results about
the autoencoder method using Keras\cite{Chollet}. In
Fig.~\ref{fig:autoencoding_2}, we show the detailed autoencoder
architecture. In the encoder, which is surrounded by the left blue
dashed rectangle, the convolutional layer 1, abbreviated as Conv1, uses
$16$ learnable filters, each of which is an array of weights, of size
$3\times 3$ with stride $1$ and zero padding of $1$.  The filter is
spatially small, but extends through the full depth of the input volume.
During the forward pass, each filter can move across the width and
height of the input volume to compute dot products between the filter
and the input.  Following the convention, stride $1$ stands for moving
the filters one pixel at a time and zero padding indicates padding the
input volume with zeros around the border. 

Similarly, Conv2, Conv3 and Conv4 use $8$ filters of size $3\times 3$
with stride $1$ and zero padding of $1$. All pooling layers are using
max pooling, which returns the maximum of the sub-region covered by the
filter, with filters of size $2\times 2$ and stride $2$. In the decoder,
surrounded by the right blue dashed rectangle, Conv5, Conv6 and Conv7
use $8$ filters of size $3\times 3$ with stride $1$ and zero padding of
$1$. Conv8 (Conv9) uses $16$ (1) filters of size $3\times 3$ with stride $1$ and
zero padding of $1$. Instead of using pooling layers, in the decoder we
apply upsampling layers, which upsamples low dimensional data to high
dimensional data through deconvolutional filters that can be learned
just like normal convolutional kernels, with filters of size $2\times
2$. The rectified linear unit (ReLU) function is activated after every
convolutional layer and the loss function used here is the binary cross
entropy loss. More details about the convolutional neural network can be
found in Ref.~\onlinecite{Wikipedia_CNNs}. The network is trained for
$100$ epochs in each case (Fig.~\ref{fig:autoencoding_3} (a), (b) or
(c)) and the training data is the raw spin configurations
$\{\mathbf{S}_i\}$ of the Ising model on a square lattice, with fixed
size $N=40\times 40$.

\end{document}